\documentclass[aps,prb,superscriptaddress,twocolumn,a4paper]{revtex4-1}
\usepackage{amsmath,empheq}
\usepackage{amssymb}
\usepackage{mathrsfs}  
\usepackage{amsfonts}
\usepackage{booktabs}
\usepackage{physics}
\usepackage{graphicx} 
\usepackage{subfigure} 
\usepackage{epsfig}
\usepackage{color}
\usepackage{epstopdf}
\usepackage{xspace}
\usepackage{rotating}
\usepackage{appendix}
\usepackage{chemformula}
\usepackage{bm}

\usepackage[colorlinks=true,linkcolor=blue,citecolor=blue,filecolor=blue,urlcolor=blue]{hyperref}
\definecolor{bondiblue}{rgb}{0.0, 0.58, 0.71}

\usepackage[utf8]{inputenc}
\usepackage[T1]{fontenc}

\begin{document}

\title{Topological strongly correlated phases in orthorhombic diamond lattice compounds}
\author{Javier Castro Luaces, Manuel Fern\'andez L\'opez, Jorge Bravo-Abad, and Jaime Merino}
\affiliation{Departamento de F\'isica Te\'orica de la Materia Condensada, Condensed Matter Physics Center (IFIMAC) and Instituto Nicol\'as Cabrera, Universidad Aut\'onoma de Madrid, Madrid 28049, Spain}
\begin{abstract}
We explore the Mott transition in orthorhombic diamond lattices relevant to (ET)Ag$_4$(CN)$_5$ molecular compounds. The non-interacting phases include nodal line, Dirac and/or Weyl semimetals depending on the strength of spin-orbit coupling and the degree of dimerization of the lattice. Based on an extension of slave-rotor mean-field theory which accounts for magnetic order, we find a transition from a semimetal to a paramagnetic Mott insulator at a critical $U_c$ which becomes N\'eel ordered at a larger Coulomb repulsion, $U_{cm}>U_{c}$. The resulting intermediate Mott phase is a $U(1)$ quantum spin liquid (QSL) consisting on spinon preserving the nodal structure of the nearby semimetallic phases. An analysis of the Green's function 
in this Mott phase shows how the zeros follow the spinon band dispersions carrying the topology while the poles 
describe the Hubbard bands. Our results are relevant to recent observations in (ET)Ag$_4$(CN)$_5$ molecular compounds in which the ambient pressure N\'eel ordered Mott insulator is gradually suppressed until semimetallic behavior arises at larger pressures. 
\end{abstract}
\date{\today}
\maketitle
\section{Introduction} 
The interplay between electron correlations and topology 
is at the forefront of research in condensed matter physics. The topological Mott insulator (TMI) as a broken symmetry ground state induced by Coulomb interaction has been proposed \cite{Zhang2008} in the context of
twisted bilayer graphene \cite{Meng2021} while in pyrochlore iridates the TMI is due to the interplay between the 
Coulomb and spin-orbit interaction (SOI) \cite{Pesin2010}. While 
topological insulators \cite{Fu2007} and Dirac semimetals \cite{KaneMele2012} have been predicted at weak Coulomb repulsion, TMIs \cite{Ashvin2009, Fiete2013,Fiete2015} and 3D quantum spin liquids \cite{Balents2007} can arise in strongly interacting frustrated diamond lattices. Mott insulators have also been observed in certain diamond lattice molecular compounds\cite{Shimizu2019}. The theoretical characterization of the topological properties across the Mott transition in these 3D semimetals is a challenging issue which may be addressed through Green's function methods \cite{Wagner2023,Sangiovanni2024}.

\begin{figure}[t!]
    \centering
\includegraphics[width=8cm,clip]{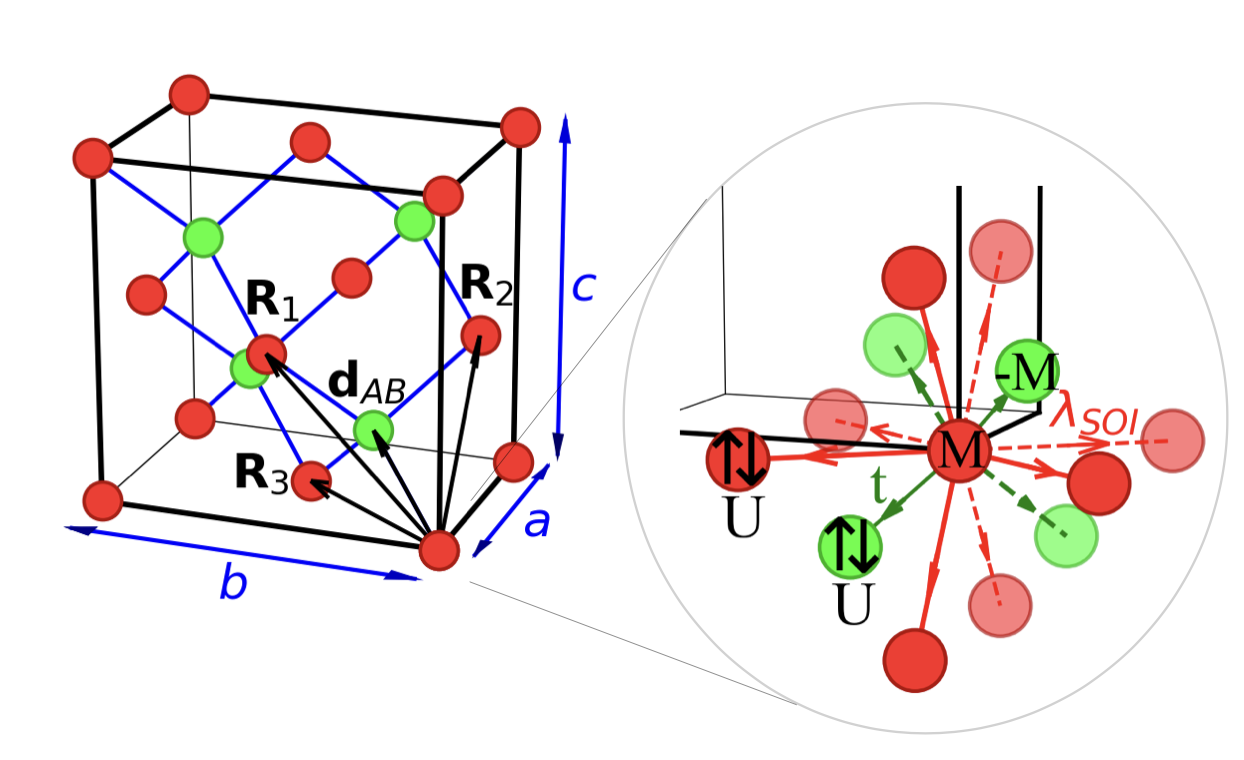}
    \caption{Crystal structure of the orthorhombic diamond lattice and model considered in our work. 
     The n.n. hopping $t$, the n.n.n. SOI, $\lambda_{SOI}$, the CDW parameter, $\pm M$,  and the onsite Hubbard $U$ in model \eqref{eq: Hubbard model} are shown as a expanded view of the chosen lattice site and its surroundings. The $\mathbf{d}_{AB}=(a/4,b/4,c/4)$ vector connects A (red circles) and B (green circles) sublattices. The coordinates of ${\bf R}_i$ vectors are given explicitly in Table \ref{tab:hoppings} of App. \ref{ap: 4}.}
\label{fig:fig1}
\end{figure}

The organic molecular compound, (ET)Ag$_4$(CN)$_5$, is an ideal platform to study the Mott transition on a diamond lattice.
The Mott insulator is suppressed under high external pressures of about 10 GPa above which semimetallic behavior has been detected \cite{Shimizu2020}. 
In these compounds, monovalent ET molecules are located at the positions of the orthorhombic diamond lattice shown in Fig. \ref{fig:fig1}. 
Hence, every ET molecule has four nearest neighbours (n.n.) belonging to the other sublattice that are located at a distance $|\mathbf{d}_{AB}|$.
The ET molecules donate an electron to Ag$_4$(CN)$_5$ anions forming honeycomb lattices surrounding the ET molecules in the $a-b$ planes, leading to half-filled bands. Band structure calculations predict a Dirac nodal line semimetal \cite{Shimizu2019,Shimizu2020} in contrast to the insulating behavior observed up to 10GPa pressure. This Mott insulator is N\'eel ordered below $T_N=102$ K and has a weak ferromagnetic component attributed to the Dzyaloshinskii-Moriya interaction implying a SOI. According to DFT \cite{Ito2020}, a n.n. hopping, $t=68.44$ meV, connects the A-B sublattices, while the effective onsite Coulomb repulsion, $U=0.71-0.78$ eV, so that $U/W \sim 1.3-1.4$ implying a Mott insulator consistent with the low pressure observations. The transition from the Mott insulator to the semimetallic behavior observed at pressures above 10 GPa 
remains theoretically unexplored. 

\begin{figure}[t!]
    \centering
\includegraphics[width=8cm,clip]{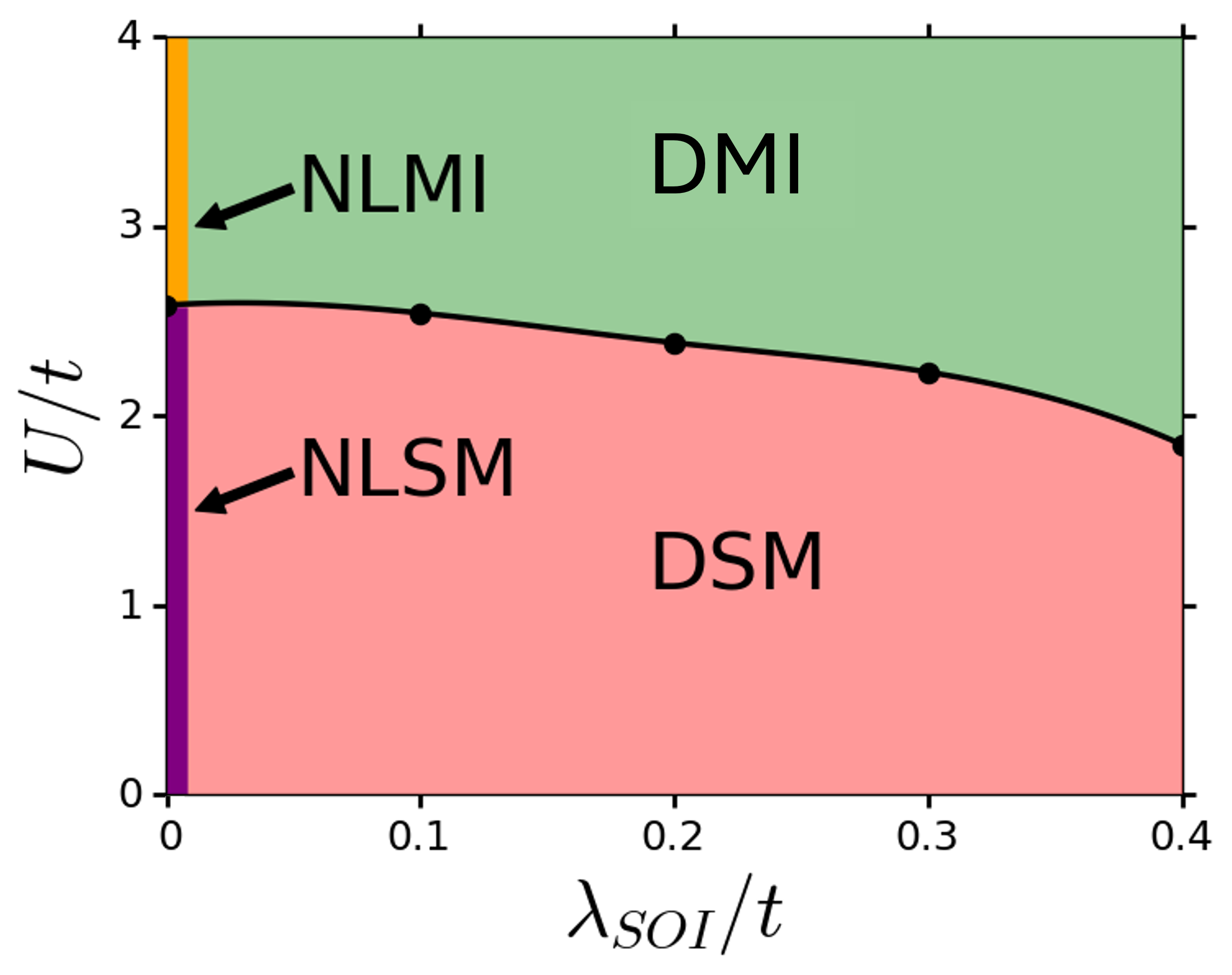} 
    \caption{U-$\lambda_{SOI}$ phase diagram of the Hubbard model on the orthorhombic diamond lattice \eqref{eq: Hubbard model} based on slave-rotor mean-field theory. With no SOI present, a direct transition from a nodal line semimetal (NLSM) to a  nodal line Mott insulator (NLMI) occurs. Any non-zero SOI induces a Dirac semimetal (DSM) at small $U$ becoming a Dirac Mott insulator (DMI) for $U$ larger than $U_c(\lambda)$, the metal-insulator transition line at $T=0$.}
\label{fig:fig2}
\end{figure}

Here, we theoretically explore the Mott transition in (ET)Ag$_4$(CN)$_5$ as a possible platform 
for TMIs in 3D. Our main results are summarized in the phase diagram of the Hubbard model on an orthorhombic diamond lattice with SOI shown in Fig. \ref{fig:fig2}. For weak Coulomb interactions and no SOI the system is a nodal line semimetal becoming a Dirac semimetal at any finite $\lambda_{SOI}$. 
A Mott metal-insulator transition occurs at $U_c$ leading to different types of topological Mott insulators depending on the strength of the SOI. While the NLMI at $\lambda_{SOI}=0$ is characterized by having nodal line spinon bands in the bulk, the DMI at
$\lambda_{SOI} \neq 0$ hosts Dirac spinons. The slave-rotor approach used in this work leads to Mott insulators in which the spin and charge degrees of freedom are fractionalized. While charge excitations are gapped, spin excitations are gapless. Since the spinons inherit the topological properties of the non-interacting semimetallic phases, the Mott insulator can be regarded a TMI.
This picture is corroborated by analyzing the Green's function across the Mott insulator transition: while the zeros follow the spinon dispersions \cite{Sangiovanni2024} characterizing the topology, the poles describe the Hubbard bands and Mott gap.

The rest of the paper is organized as follows. In Sec. \ref{sec: model} we introduce a Hubbard model on an orthorhombic diamond lattice to explore the Mott transition. In Sec. \ref{sec: TSM} we analyze the various non-interacting semimetallic phases arising in the model depending on the various ingredients such as the SOI or CDW order parameter. In Sec. \ref{sec: strongcoupling} we discuss the strongly interacting limit of the Hubbard model introduced in Sec. \ref{sec: model}. Sec. \ref{sec: SRMFT} is devoted to the Mott transition and slave rotor mean-field theory. The connection between spinon bands and Green's function zeros is discussed in Sec. \ref{sec:zeros}. In Sec. \ref{sec: obs} we discuss our results in the context of experimental observations in (ET)Ag$_4$(CN)$_5$ molecular compounds. In Sec. \ref{sec: conclusions} we summarize our main results and discuss possible extensions of our work beyond slave rotor mean-field theory. 

\begin{figure*}[t!]
    \centering
\includegraphics[width=0.9\linewidth]{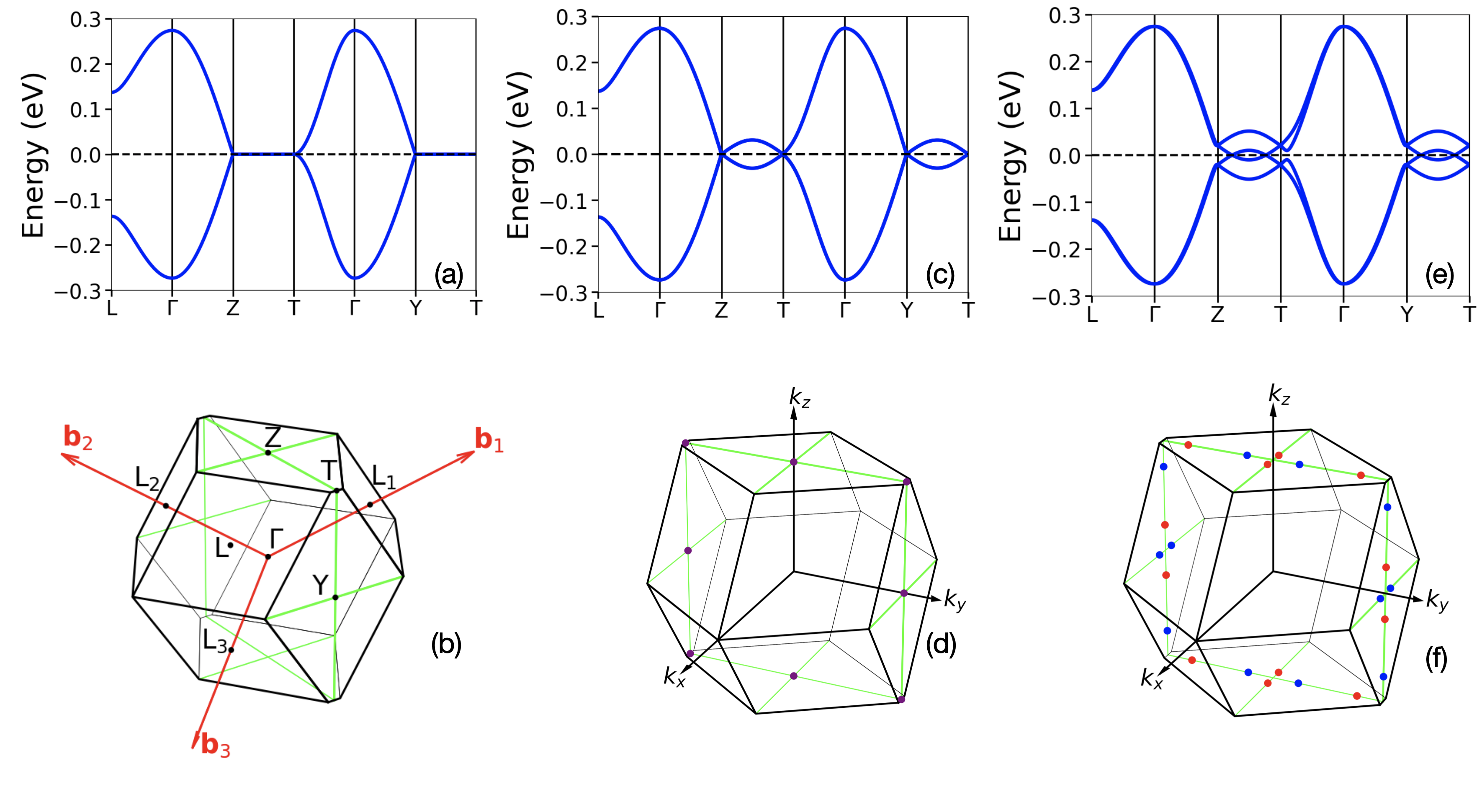}
   \caption{Electronic band structure of semimetals arising in model 
   \eqref{eq: Hubbard model} at $U=0$. (a) NLSM arising for $\lambda_{SOI}=M=0$
   displaying in (b) the corresponding Dirac nodal lines (green lines) 
   in the first Brillouin zone. (c) DSM for $\lambda_{SOI}=0.2t, M=0$ showing the Dirac nodes (purple dots) in (d). (e) Weyl semimetal for $\lambda_{SOI}=0.2t$ and $M=1.5 \lambda_{SOI}$ displaying its associated Weyl nodes in (f). The red vectors in (b): $\mathbf{b}_1=2\pi(-1/a, 1/b,1/c)$, $\mathbf{b}_2=2\pi(1/a, -1/b,1/c)$ and $\mathbf{b}_3=2\pi(1/a,1/b,-1/c)$ are the vectors of the primitive unit cell in reciprocal space and $\Gamma=(0,0,0)$, $Z=(0,0,\frac{2\pi}{c})$, $Y=(0,\frac{2\pi}{b},0)$, $T=(0,\frac{2\pi}{b},\frac{2\pi}{c})$, $L=(\frac{\pi}{a},\frac{\pi}{b},\frac{\pi}{c})$, $L_1=(-\frac{\pi}{a},\frac{\pi}{b},\frac{\pi}{c})$, $L_2=(\frac{\pi}{a},-\frac{\pi}{b},\frac{\pi}{c})$, $L_3=(\frac{\pi}{a},\frac{\pi}{b},-\frac{\pi}{c})$ are the eight TRIM points. The Dirac points (purple dots) in (d) split into Weyl points with positive (red dot) and negative (blue dot) chiralities in (f). We take $\gamma=1$ and  $t=-68.442$ meV in all band structures.}
   \label{fig:bandstructure}
\end{figure*}

\section{Hubbard model on orthorhombic diamond lattice}
\label{sec: model}
We analyze the Mott transition on the orthorhombic diamond
lattice of Fig. \ref{fig:fig1} based on a generalized Hubbard model extended to include an SOI and an 
alternating charge density wave (CDW) potential. Thus, the 
complete model reads:
\begin{equation}
\mathcal{H}=\mathcal{H}_{NLSM}+\mathcal{H}_{SOI} + \mathcal{H}_{CDW}+  \mathcal{H}_{U},
    \label{eq: Hubbard model}
\end{equation}
where:
\begin{eqnarray}
  \mathcal{H}_{NLSM} &=& \sum_{\langle i,j\rangle} \sum_\alpha t_{ij}
  c_{i\alpha}^\dagger c_{j\alpha},
  \nonumber \\
 \mathcal{H}_{SOI} &=& i\lambda_{SOI} \sum_{\langle\langle i,j\rangle \rangle}\sum_{\alpha,\beta}c_{i\alpha}^\dagger \mathbf{\tau}_{\alpha\beta}\cdot\frac{\mathbf{d}_{il}\times 
 \mathbf{d}_{lj}}{|\mathbf{d}_{il}\times \mathbf{d}_{lj}|}c_{j\beta},
\nonumber \\
\mathcal{H}_U  &=& \frac{U}{2}\sum_i (n_i-1)^2,
\nonumber \\
\mathcal{H}_{CDW} &=& \sum_{i,\sigma}M_i n_{i\sigma}. 
\end{eqnarray}
Since the lattice is bipartite (see Fig. \ref{fig:fig1}), $t_{ij}=t$, is the hopping between the two sublattices located in different unit cells whereas, $t_{ij}= \gamma t$, 
is the hopping between two sublattices in the same unit cell. If  $\gamma \neq 0$ spatial isotropy is broken which could be achieved by applying uniaxial pressure along the $[111]$ direction of the (ET)Ag$_4$(CN)$_5$ compound. We generally take $t=-68.442\ \mathrm{meV}$, and $\gamma=1$ as obtained from DFT calculations\cite{Shimizu2019} on (ET)Ag$_4$(CN)$_5$ discussing the $\gamma \neq 1$ anisotropy whenever relevant.  $\mathcal{H}_{SOI}$ is a Fu-Kane-Mele spin-orbit contribution \cite{Fu2007,Vanderbilt2018} and $\mathcal{H}_{CDW}$ an alternating charge density wave (CDW) potential. Here $\alpha=\uparrow,\downarrow$ labels the two spin degrees of freedom, $\mathbf{\tau}=(\tau^1,\tau^2,\tau^3)$ is the vector of Pauli matrices acting on spin space. $\mathbf{d}_{il}$ and $\mathbf{d}_{lj}$ are bond vectors connecting n.n. sites adding up to bonds between n.n.n. sites on the diamond lattice (see Fig. \ref{fig:fig1} for the lattice geometry). The SOI term is hermitian, since $\mathbf{d}_{il}=-\mathbf{d}_{li}$, $ \mathbf{d}_{lj}=-\mathbf{d}_{jl}$ and $\mathbf{\tau}_{\alpha\beta}=\mathbf{\tau}^*_{\beta\alpha}$, preserving 
$\mathcal{T}$ symmetry. We assume an alternating potential $M_i=M,-M$ in sublattices A and B, respectively, with $M \geq 0$ in $\mathcal{H}_{CDW}$. This term breaks $\mathcal{P}$-symmetry but preserves $\mathcal{T}$-reversal symmetry. Finally, $\mathcal{H}_U$ is a standard 
onsite Hubbard Coulomb repulsion.


We explore in the following our model \eqref{eq: Hubbard model} in different parameter regimes. As shown below, in the non-interacting limit,  $U \rightarrow 0$
we can have a nodal line, a Dirac or a Weyl semimetal depending on $\lambda_{SOI}$ and $M$. At strong coupling, the model can be mapped onto a FM $x-y$ model with an AFM Ising interaction in the $z$-direction. The Mott transition at intermediate $U$ is explored based on slave rotor mean-field theory (SRMFT).

\section{Topological semimetals}
\label{sec: TSM}
At $U=0$, we can neglect $\mathcal{H}_U$ and different band structures 
arise depending on the terms kept in the non-interacting hamiltonian. We 
first consider three different semimetals with isotropic hoppings, 
$\gamma=1$, but different parameters: 
(i) $\lambda_{SOI}=M=0$, (ii) $\lambda_{SOI} \neq 0$, 
$M=0$, (iii) $\lambda_{SOI} \neq 0 $, $M \neq 0$. We finally consider the possibility of a topological insulator with (iv) $\gamma \neq 1$ and $\lambda_{SOI} \neq 0 $. Case (ii) corresponds to a 3D Fu-Kane-Mele type of model\cite{Fu2007}. We discuss these three cases paying special attention to their associated topological properties. 

{\em Nodal line semimetal, $\lambda_{SOI}=M=0$. }
We first consider the simplest non-interacting model, a n. n. tight-binding model on the diamond lattice:
\begin{equation}
    \mathcal{H}_{NLSM}=\sum_{\langle i,j\rangle } t_{ij}\sum_\alpha c_{i\alpha}^\dagger c_{j\alpha},
    \label{eq: NLSM}
\end{equation}
In reciprocal space, the model can be more simply expressed in terms of Pauli matrices as:
\begin{equation}
    \mathcal{H}_{NLSM}(\mathbf{k})=d_1(\mathbf{k})\sigma^1+d_2(\mathbf{k})\sigma^2,\label{eq:NLSM_r}
\end{equation}
where the Pauli matrices are now denoted by $\sigma^a$ with $a=1,2,3$ (corresponding to $x,y,z$ components) and act on the sublattice space 
$\mathfrak{L}$ ($\mathfrak{L}=\{A,B\}$), with:
\begin{equation}   
d_1(\mathbf{k})=t (\gamma+\text{cos}(\mathbf{k\cdot R}_1)+\text{cos}(\mathbf{k\cdot R}_2)+\text{cos}(\mathbf{k\cdot R}_3)),\label{eq: d1 NLSM hamiltonian}
\end{equation}
\begin{equation}
d_2(\mathbf{k})=t (\text{sin}(\mathbf{k\cdot R}_1)+\text{sin}(\mathbf{k\cdot R}_2)+\text{sin}(\mathbf{k\cdot R}_3)).\label{eq: d2 NLSM hamiltonian}
\end{equation}
The band structure associated with $\mathcal{H}_{NLSM}(\mathbf{k})$ is shown in Fig. \ref{fig:bandstructure} (a) for $\gamma=1$. It is worth noting the band degeneracies arise along the $Z-T$ and $Y-T$ segments of the Fermi energy, $\epsilon_F=0$, of the half-filled system. A simple analysis of \eqref{eq:NLSM_r} shows that the 
dispersion relation can be expressed as:
\begin{equation}
\epsilon_\pm(\mathbf{k})=\pm\sqrt{d_1^2(\mathbf{k})+d_2^2(\mathbf{k})},\label{eq: Ener NLSM}
\end{equation}
so that the band degeneracy at ${\bf k}$ points must satisfy the condition:
\begin{equation}
    d_1(\mathbf{k})=d_2(\mathbf{k})=0,
    \label{eq:deg 0}
\end{equation}
at $\epsilon_F=0$. The dimension of this degeneracy is found to be equal to $D-\delta_{CL}$, with $D$ being the dimensionality of the system ($D=3$) and $\delta_{CL}$ the codimension of the node.
This codimension refers to the number of equations a $\mathbf{k}$-point has to verify for 
accommodating a degeneracy which from \eqref{eq:deg 0} we can see that
$\delta_{CL}=2$. Moreover, a discussion in \cite{Sigrist2017} shows that in two-band systems the codimension of the nodes is found to be equal to the minimum number of different Pauli matrices necessary for expressing the hamiltonian \eqref{eq: NLSM}. This arises from the fact that for every Pauli matrix that appears, an equation of the form $d_a(\mathbf{k})=0$ can be considered as a new condition 
on the $\mathbf{k}$-point to display a band degeneracy. Here, $d_a(\mathbf{k})$ is the coefficient associated with a
particular Pauli matrix $\sigma^a$.


Since $D-\delta_{CL}=1$ band degeneracies must form one-dimensional lines in $\mathbf{k}$-space, which are denoted as nodal lines or nodal loops if they are closed. The set of $\mathbf{k}$ satisfying \eqref{eq:deg 0} lead to 
three closed mutually perpendicular rectangular nodal lines centered at the $\Gamma$-point as shown in Fig. \ref{fig:bandstructure} (b). 

If hopping terms up to four n.n. are considered, the Hamiltonian in reciprocal space would read:
\begin{equation}    
\mathcal{H}_{NLSM}(\mathbf{k})=f_0(\mathbf{k})\sigma^0+f_1(\mathbf{k})\sigma^1+f_2(\mathbf{k})\sigma^2,\label{eq: NLSM4}
\end{equation}
with $\sigma^0$ being the identity matrix. The dispersion relation becomes: 
\begin{equation}
\epsilon_\pm(\mathbf{k})=f_0(\mathbf{k})\pm\sqrt{f_1^2(\mathbf{k})+f_2^2(\mathbf{k})},
\end{equation}
where expressions for $f_0(\mathbf{k})$, $f_1(\mathbf{k})$
and $f_2(\mathbf{k})$ are given in App. \ref{ap: 4}. Since $\sigma^0$ is not a Pauli matrix, it is irrelevant for determining the node codimension and so
remaining $\delta_{CL}=2$ also in this case. Although this system is also characterized by the presence of nodal lines the Fermi surface consists of electron and hole pockets\cite{Shimizu2019,Ito2020} (see App. \ref{ap: 4}).

We now consider the topological properties of the NLSM described by a 3D
model of the type \eqref{eq: NLSM} with codimension, $\delta_{CL}=2$. Due to their codimension, $\delta_{CL}$=2, \cite{Sigrist2017} the only $p$-spheres $S^p$  (spheres of dimension $p$) that can wrap the nodal loops accommodating a topological charge are those with $p= 1$ (loop), 2 (sphere). Hence, nodal loops are characterized by two independent topological indices $\zeta_p$, which belong to the $\mathbb{Z}_2$ homotopy group \cite{PhysRevB.92.081201} and give information on the way in which the nodal loops evolve when perturbing the Hamiltonian while preserving its $\mathcal{P}$, $\mathcal{T}$ and SU(2) symmetries. 

The topological index $\zeta_1$ is simply the Berry phase over a ring 
$S^1$ that links with the nodal loop,
\begin{equation}
    \zeta_1=\oint_{S^1}\mathbf{\mathcal{A}}(\mathbf{k})d\mathbf{k}\ \ \ \text{mod}\ 2\pi,
    \label{eq:berry}
\end{equation}
where
\begin{equation}
\displaystyle
    \mathbf{\mathcal{A}}(\mathbf{k})=i\sum_{\mu\in occ.}\bra{u_\mu(\mathbf{k})}\partial_{\mathbf{k}}\ket{u_\mu(\mathbf{k})},
\end{equation}
is the Berry connection and $n$ a band index. Since in this case we only have one occupied band, described by the Bloch eigenstate, $\ket{u_-}$, with eigenenergy $\epsilon_-(\mathbf{k})$ given in \eqref{eq: Ener NLSM}. This state takes the form:
\begin{equation}
\ket{u_{-}(\mathbf{k})}=\frac{1}{\sqrt{2}}\begin{pmatrix}
    \frac{\epsilon_-(\mathbf{k})}{d_1(\mathbf{k})+id_2(\mathbf{k})}\\
    1\\
\end{pmatrix}
\end{equation}
in our chosen particular gauge. Thus,
\begin{equation}
    \mathcal{A}(\mathbf{k})=\frac{d_2(\mathbf{k})\partial_{\mathbf{k}}d_1(\mathbf{k})-d_1(\mathbf{k})\partial_{\mathbf{k}}d_2(\mathbf{k})}{2\epsilon_-({\bf k})^2}.
    \end{equation}
As shown in \cite{PhysRevB.97.161113}, a null value of $\zeta_1$ indicates that the degeneracy is accidental and removable by any small perturbation
preserving the symmetries of the hamiltonian. On the other hand, a non-zero value of the Berry phase means that the nodal loop is protected by SU(2) and $\mathcal{PT}$ symmetries. 

The robustness of the nodal loops against a small perturbation
preserving all hamiltonian symmetries can be analyzed by modifying $\gamma$ in $d_1(\mathbf{k})$ of \eqref{eq: d1 NLSM hamiltonian} with $\gamma\in \mathbb{R}$ which dimerizes the hoppings along the $[111]$ direction. Note that the present NLSM described by hamiltonian \eqref{eq: NLSM} falls in the AIII class 
according to the classification scheme of Ref. \onlinecite{Sigrist2017} (see Table II
in this reference).
\begin{figure}[b]
    \centering
    \includegraphics[width=0.8\linewidth]{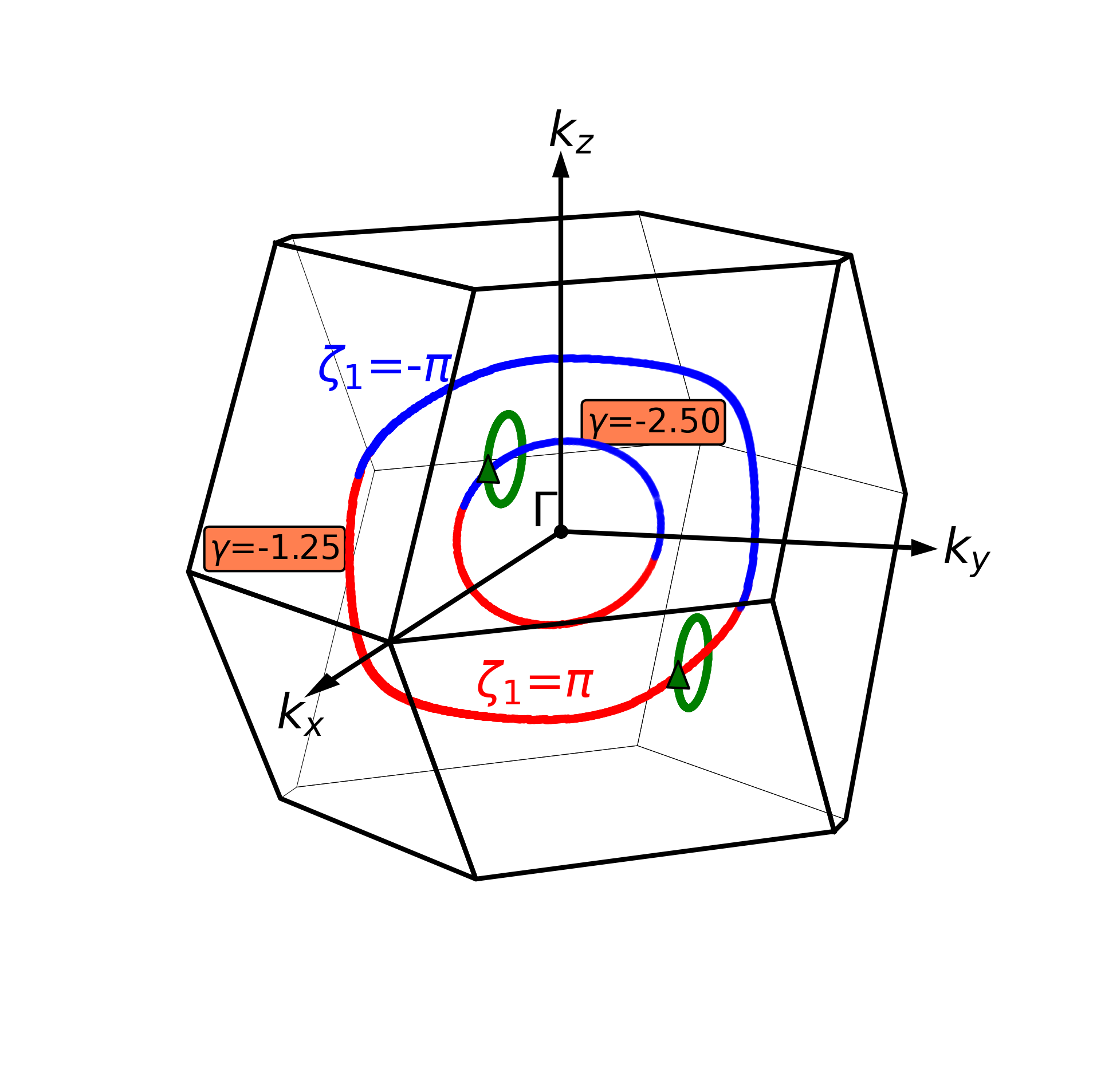}
    \caption{Dependence of nodal lines of the NLSM on lattice dimerization $\gamma$ along the $[111]$ direction. The Berry phases, $\zeta_1$, along the nodal lines calculated from the Berry flux traversing the green interlinked rings, are shown. Note that, despite the perspective, the nodal lines are not necessarily in the same plane.}
    \label{fig:fig4}
\end{figure}

In Fig. \ref{fig:fig4}, the dependence of nodal loops on $\gamma <0$ is shown. As $\gamma$ decreases, the nodal loop contracts towards the $\Gamma$-point, until the critical $\gamma=-3$ is reached, at which the nodal line becomes 
a single point localized at $\Gamma$. For $\gamma<-3$, the nodal loop disappears and the system becomes gapped. A similar situation occurs for $\gamma >0$ (not shown). The only differences are that the TRIM to which the loop contracts is now L, and that the critical value at which the loop disappears is $\gamma=3$ \cite{TFM}. The Berry phase, $\zeta_1$ along the nodal loop is obtained from Eq. \eqref{eq:berry} by integrating over the green interlinked rings shown in Fig. \ref{fig:fig4}. Since half of the nodal line has $\zeta_1=\pi$ and the other half has: $\zeta_1=-\pi$, the
total Berry phase inside the FBZ is zero. Hence, the second topological index of the nodal loop, $\zeta_2=0$, indicating that it cannot be considered a $\mathbb{Z}_2$ charge monopole. 
This is consistent with the fact that similarly to Weyl nodes
(except that Weyl nodes can also have a $\mathbb{Z}$-charge of -1), charged nodal lines can only be created and annihilated in pairs but cannot shrunk to a point becoming gapped when continuously varying Hamiltonian parameters without breaking 
the symmetries as happens in our present case for $|\gamma| >3$.

In the studied NLSM, the bulk-boundary correspondence guarantees the presence of in-gap states within the projected Surface Brillouin Zone of a system characterized by a nodal loop with quantized Berry phases in its bulk.\cite{Chan_2016} These surface states, confined to the projection of the nodal loop onto the surface, exhibit a nearly flat dispersion reminiscent of a drumhead, hence the term "drumhead states". This flat dispersion results in an exceptionally high density of states and significant correlation effects,\cite{Drumhead} positioning these systems as promising platforms for exotic electronic phenomena, such as high-temperature superconductivity\cite{PhysRevB.83.220503} and the emergence of Majorana fermions.\cite{PhysRevB.103.035133}
Beyond theoretical interest, experimental studies of these systems have accelerated in recent years, notably with the detection of drumhead surface states in nodal-line semimetallic materials via ARPES techniques.\cite{Muechler2020,Hosen2020}

In summary, the nodal lines of the NLSM occurring for $|\gamma| < 3$ have non-zero Berry phases $\zeta_1 = \pm \pi$ along the nodal line but are uncharged, $\zeta_2=0$. These type of nodal lines are one-dimensional analogues of Dirac nodes so they are more specifically denoted Dirac nodal lines.

{\em Dirac semimetal: $\gamma=1$, $\lambda_{SOI} \neq 0$. } We now consider our non-interacting model which includes SOI effects on the NLSM:
\begin{align}
    \mathcal{H}_{DSM}&=\mathcal{H}_{NLSM}+\mathcal{H}_{SOI}.\notag \\
    \label{eq:Dirac H}
\end{align}
Since SU(2) symmetry is broken, $\mathcal{H}_{DSM}$, is no longer 
$2\times2$ but $4\times4$. Therefore, $\mathcal{H}_{DSM}$ cannot be expressed 
through Pauli matrices but rather in terms of tensor products 
among them:
\begin{equation}
\mathcal{H}_{DSM}(\mathbf{k})=\sum_{\alpha,\beta}^3 g_{\alpha\beta}(\mathbf{k})\sigma^\alpha\otimes\tau^\beta,
\end{equation}
with $g_{\alpha\beta}(\mathbf{k}) \in \mathbb{R} $ and 
$\alpha=0,1,2,3$. Pauli matrices $\sigma^\alpha$ acts in the lattice 
subspace $\mathfrak{L}\ (\mathfrak{L}=\{A,B\})$ whereas Pauli matrices
$\tau^\beta$ acts on the spin subspace $\mathfrak{G} \ (\mathfrak{G}=\{\uparrow,\downarrow\})$. The different tensor products between Pauli matrices give rise to 16 Dirac matrices, $\Gamma^r$.

Our Hamiltonian preserves space-time reversal symmetry, {\it i.e.} $\mathcal{S}\mathcal{H}_{DSM}(\mathbf{k})\mathcal{S}^{-1}=\mathcal{H}_{DSM}(\mathbf{k})$ with $\mathcal{S}=\mathcal{P}\mathcal{T}$, thus it can be expanded in terms of matrices that 
commute with $\mathcal{S}$. We will search for these among the 16 Dirac matrices. Since the inversion operator swaps $A\leftrightarrow B$, while leaving spin unchanged, it can be represented as $\mathcal{P}=\sigma^1\otimes\tau^0$. The time-reversal operator for spin-1/2 particles is represented as $\mathcal{T}=-i(\sigma^0\otimes\tau^2)\mathcal{K}$ in the convention of a $\pi$ rotation around the spin $y$-axis, with $\mathcal{K}$ indicating complex conjugation. Both symmetry operations also reverse $\mathbf{k}\rightarrow -\mathbf{k}$ when acting on functions of such variable, therefore, their composition is evidently the identity over $\mathbf{k}$-space. Thereby, $\mathcal{S}=\mathcal{P}\mathcal{T}=-i(\sigma^1\otimes\tau^2)\mathcal{K}$ and $\mathcal{S}^{-1}=i\mathcal{K}(\sigma^1\otimes\tau^2)$, where the commutation of a generic element 
$\sigma^\alpha\otimes\tau^\beta$ with this operator is reduced to the 
condition   $(\sigma^\alpha\otimes\tau^\beta)^*=(\sigma^1\sigma^\alpha\sigma^1)\otimes(\tau^2\tau^\beta\tau^2)$. From the 16 different Dirac matrices only 6 of them 
fulfill the previous condition, the so called $\mathcal{PT}$-even Dirac matrices:

\begin{align}
    \Gamma^0=\sigma^0\otimes\tau^0,\ \Gamma^1=\sigma^1\otimes\tau^0,\ \Gamma^2=\sigma^2\otimes\tau^0,\notag\\
    \Gamma^3=\sigma^3\otimes\tau^1,\ \Gamma^4=\sigma^3\otimes\tau^2,\ \Gamma^5=\sigma^3\otimes\tau^3. \label{eq: Dirac matrices}
\end{align}
Thus, the Hamiltonian can be expressed in the reciprocal space in terms solely of these 6 matrices. In this way, we have managed to reduce the dimension of the representation from 16 to 6, which in fact are actually 5 matrices, since $\Gamma^0$ does not contribute up to the n.n. hoppings:
\begin{equation}
    \mathcal{H}_{DSM}(\mathbf{k})=\sum_{r=1}^5d_r(\mathbf{k})\Gamma^r.\label{eq:DSM}
\end{equation}
The different $d_r(\mathbf{k})$ associated to this Hamiltonian can be seen in App. \ref{ap: DSM}.

Similarly to the two-band case, the dispersion relation of this Hamiltonian 
reads:
\begin{equation}
\epsilon_\pm(\mathbf{k})=\pm  \sqrt{d_1^2(\mathbf{k})+d_2^2(\mathbf{k})+d_3^2(\mathbf{k})+d_4^2(\mathbf{k})+d_5^2(\mathbf{k})}.\label{eq: Dirac energy}
\end{equation}
Under the presence of $\mathcal{PT}$ symmetry, Kramers degeneracy always enforces a two-fold spin degeneracy. This is why, despite our hamiltonian being $4\times4$, the band structure consists of only two bands which are two-fold degenerate.


As stated above, the node codimension is equal to the minimum number of Dirac matrices $\Gamma^r$ needed for expressing the hamiltonian. Therefore, $\delta_{CL}=5>D$, and a four-fold degeneracy should be ruled out under the effect of this hamiltonian. This is true for all the different points of $\mathbf{k}$-space, except for those belonging to the TRIM. Recalling that $\Gamma^1=\mathcal{P}$, we have that

\begin{equation}
\mathcal{H}_{DSM}(-\mathbf{k})=\mathcal{P}\mathcal{H}_{DSM}(\mathbf{k})\mathcal{P}^{-1}=d_1(-\mathbf{k})\Gamma^1-\sum_{i=2}^5d_i(-\mathbf{k})\Gamma^i,
\end{equation}
where we have taken advantage of the fact that Dirac matrices satisfy $\{\Gamma^r,\Gamma^s\}=2\delta_{rs}\Gamma^0,\ \forall r,s=1,...,16$ (Euclidean Clifford algebra). Since at a TRIM $\mathbf{k}=-\mathbf{k}$, we must have:

\begin{equation}
 d_2(\mathbf{k})= d_3(\mathbf{k})= d_4(\mathbf{k})= d_5(\mathbf{k})=0\ \text{if $\mathbf{k}$ is a TRIM},           
\end{equation}
for any function $d_i(\mathbf{k}) \in \mathbb{R} $. Therefore, the codimension at a TRIM reduces to $\delta_{CL}=1<D$,  since it only has to verify the equation $d_1(\mathbf{k})=0$ for accommodating a four-fold degeneracy. This makes TRIM points very prone to hosting degeneracies among all bands in these systems.

The band structure associated to the SOI Hamiltonian \eqref{eq:Dirac H} is shown in Fig. \ref{fig:bandstructure} (c). Apart from the two-fold spin degeneracy found throughout the entire $\mathbf{k}$-space, the band structure features four-fold degeneracies at the TRIM points $Z$, $T$, and $Y$. Notice that since the TRIM are the only points in reciprocal space at which $\delta_{CL}=1$, the four-fold degeneracies appear as disconnected points despite $D-\delta_{CL}=2$ (surface) as shown in Fig. \ref{fig:bandstructure}. Since the dispersion at these four-fold degenerate points is linear they can be regarded as Dirac cones characterizing a Dirac semimetal.

{\it Weyl semimetal: $\gamma=1$, $\lambda_{SOI}, M \neq 0$.}
We consider the effect of a CDW potential on the DSM through the model:
\begin{equation}
\mathcal{H}_{WSM}=\mathcal{H}_{NLSM}+\mathcal{H}_{SOI}+\mathcal{H}_{CDW}.
\end{equation}
The CDW term leaves the spin subspace intact but acts with opposite signs on $A$ ($+M$) or $B$ ($-M$) sublattices. Based on this, it is straightforward to see how the Dirac matrix associated to $\mathcal{H}_{CDW}({\bf k})$ is $\Gamma^6=\sigma^3\otimes\tau^0$. Therefore, our new Hamiltonian in the Dirac matrix representation reads:
\begin{equation}
    \mathcal{H}_{WSM}(\mathbf{k})=\mathcal{H}_{DSM}(\mathbf{k})+\mathcal{H}_{CDW}(\mathbf{k})=\sum_{i=1}^5d_i(\mathbf{k})\Gamma^i+M\Gamma^6,\label{eq: Weyl Hamiltonian}
\end{equation}
which is readily diagonalized leading to four bands:
\begin{align}
   \epsilon_\pm^C(\mathbf{k})=\sqrt{\sum_{i=1}^2d_i^2(\mathbf{k})+\left[M\pm\sqrt{\sum_{i=3}^5d_i^2(\mathbf{k})}\right]^2},\notag \\
   \epsilon_\pm^V(\mathbf{k})=-\sqrt{\sum_{i=1}^2d_i^2(\mathbf{k})+\left[M\pm\sqrt{\sum_{i=3}^5d_i^2(\mathbf{k})}\right]^2}.\label{eq: Weyl dispersion}
\end{align}

Here, the superindices $V$ and $C$ denote valence and conduction bands, respectively. Hence, by breaking the $\mathcal{P}$ symmetry, we have lifted Kramer's degeneracy present at $M=0$. Still many degeneracies can occur among pairs of bands at specific ${\bf k}$-points of the FBZ. Despite this we focus on the degeneracies occurring at the Fermi level, $\epsilon_F=0$, since these characterize the low energy electronic properties of the system. From \eqref{eq: Weyl dispersion} it is easy to see that a degeneracy with $\epsilon(\mathbf{k})=0$ can only occur between $\epsilon_-^C(\mathbf{k})$ and $\epsilon_-^V(\mathbf{k})$. 
The set of $\mathbf{k}$-points at which band degeneracies occur are defined by the conditions:
\begin{equation}
    d_1(\mathbf{k})=d_2(\mathbf{k})=M^2-d^2_3(\mathbf{k})-d^2_4(\mathbf{k})-d^2_5(\mathbf{k})=0,
\end{equation}
where the codimension of the band degeneracy is $\delta_{CL}=3$
leading to Weyl nodes in the 3D ${\bf k}$-space. 

Weyl nodes are characterized by their $\mathbb{Z}$-charge or
chirality:
   $ \chi=-\frac{1}{2\pi}\oint_{S^2}\mathbf{\Omega}(\mathbf{k})d\mathbf{S}=-C,$
with $S^2$ a spherical surface wrapping the Weyl node, $\mathbf{\Omega}(\mathbf{k})=\mathbf{\nabla}\times\mathbf{\mathcal{A}}(\mathbf{k})$ is the Berry curvature and $C$ the Chern number.  
The computation of $\chi$ can be greatly simplified by noting that the Weyl nodes can be described through a $2 \times 2$ hamiltonian:
\begin{equation}
  \mathcal{H}_{eff}(\mathbf{k})=d_1(\mathbf{k})\sigma^1+d_2(\mathbf{k})\sigma^2+\left[M+\sqrt{\sum_{i=3}^5d_i(\mathbf{k})}\right]\sigma^3,  
\end{equation}
as they only involve two bands: $\epsilon_-^C$ and $\epsilon_+^V$. Following \cite{Vanderbilt2018,Goikoetxea2020}
the chiralities of the Weyl nodes described by by a hamiltonian of the 
form $\mathcal{H}_{2\times2}(\mathbf{k})=\sum_{a=1}^3f_a(\mathbf{k})\sigma^a$ can be expressed as:
\begin{equation}
    \chi=\text{sgn} \left[ det \left( \partial f_b(\mathbf{k}_W)\over\partial  {k_a} \right) \right],
\end{equation}
with $ det \left(\partial f_b (\mathbf{k}_W) \over\partial  {k_a} \right) $, the determinant of the $3\times3$ Jacobian matrix evaluated at the Weyl point located at $\mathbf{k}_W$. 

As $M$ is increased each Dirac cone at the $Z$, $T$ and $Y$ TRIM points split into two Weyl nodes of opposite chiralities, $\chi =\pm 1$, as it should. The band structure at $M=1.5 \lambda_{SOI}, \lambda_{SOI}=0.2t$ is shown in Fig. \ref{fig:bandstructure} (e) which hosts the 24 Weyl nodes shown in Fig. \ref{fig:bandstructure} (f). As expected from the Nielsen-Ninomiya ``fermion doubling'' theorem\cite{1983PhLB..130..389N}, Weyl nodes occur in pairs of opposite chiralities so the system
has zero net chirality. At a critical $M=M_c$, Weyl nodes of opposite $\chi$ annihilate finally becoming a trivial band insulator at larger $M>M_c$.

\begin{figure}[h]
    \centering
    \includegraphics[width=0.95\linewidth]{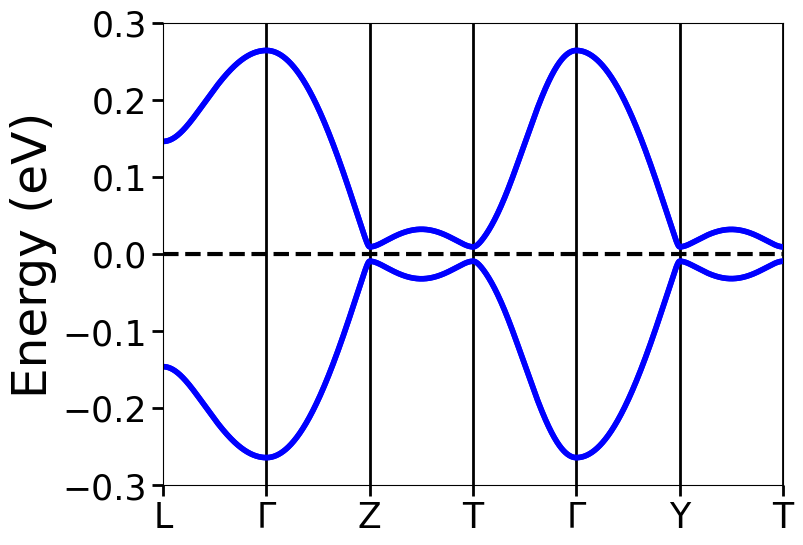}
    \caption{Band structure of the Kane-Fu-Mele type model \eqref{eq:Dirac H} on the dimerized ortorhombic diamond lattice. The dimerization along the $[111]$ direction has caused our system to become now an insulator. We have fixed $\gamma=3.5$ and $\lambda_{SOI}=0.2t$ with $t=-68.442$ meV.}    \label{fig:fig5}
\end{figure}

{\it Topological insulator: $\gamma \neq 1, \lambda_{SOI} \neq 0$.} 
Finally, we consider the possibility of stabilizing, at weak coupling,
a topological insulator induced by SOI. As discussed above, model \eqref{eq:Dirac H} leads to Dirac cones in the band structure. By introducing  
a distortion in the $[111]$ bonds by taking $\gamma \neq 1$, in the presence 
of $SOI$ we can open a gap in the system leading to a topological insulator as shown in \ref{fig:fig5}. The topology of an insulator where $\mathcal{T}$ symmetry is preserved is described by four $\mathbb{Z}_2$ independent indices \cite{End}, usually displayed as $(\nu_0;\nu_1\nu_2\nu_3)$ which can take odd and even values ($\nu_i=0,1\ \text{mod 2}$). They are divided between strong ($\nu_0$) and weak ($\nu_{j=1,2,3}$) indices, with the strong one being the most important. An insulator with an odd value of $\nu_0$ is classified as a strong topological insulator (STI), meanwhile if it presents an even value of $\nu_0$ it is said to either be a weak topological insulator (WTI) or to be topologically trivial. The difference between weak and trivial topology is given by the three weak indices. An insulator with an even $\nu_0$ but at least one $\nu_j$ odd is said to be topologically weak whereas an insulator described by the index set ($0;000$) is topologically trivial. 

If, in addition to being $\mathcal{T}$-symmetric, a system presents $\mathcal{P}$ symmetry, like in the present case, the computation of these indices is greatly simplified since they depend only on the parity of each pair of Kramers degenerate occupied bands at the eight TRIM $\Gamma_i$ in the FBZ \cite{PhysRevB.76.045302}. These special points can be expressed in terms of the primitive reciprocal lattice vectors as $\Gamma_{i=(n_1n_2n_3)}=(n_1\mathbf{b}_1+n_2\mathbf{b}_2+n_3\mathbf{b}_3)/2$. As in the Fu-Kane-Mele model there is a single pair of Kramer degenerate occupied bands, these topological indices can be expressed as:
\begin{equation}
    (-1)^{\nu_0}=\prod_{n_k=0,1}\xi(\Gamma_{i=(n_1n_2n_3)}),
\end{equation}

\begin{equation}
    (-1)^{\nu_{k=1,2,3}}=
    \prod_{n_k=1, n_{j\neq k}=0,1}\xi(\Gamma_{i=(n_1n_2n_3)}),
\end{equation}
where $\xi(\Gamma_i)$ is the eigenvalue associated to the parity operator when measured over the pair of Kramer degenerate occupied bands at a $\Gamma_i$ TRIM. Note that the four points describing each index $\nu_{j=1,2,3}$ all lay on the same plane. Moreover, in the same notation used previously 
for the TRIM, the weak indices read:
\begin{align}
    (-1)^{\nu_1}=\xi(L)\xi(L_1)\xi(Z)\xi(Y),\notag \\
    (-1)^{\nu_2}=\xi(L)\xi(L_2)\xi(Z)\xi(T),\notag \\
    (-1)^{\nu_3}=\xi(L)\xi(L_3)\xi(Z)\xi(T).
\end{align}
In the present model we have $\Gamma^1=\sigma^x\otimes\tau^0=\mathcal{P}$ and
at the TRIM the Hamiltonian reduces to $\mathcal{H}(\mathbf{\Gamma}_i)=d_1(\mathbf{\Gamma}_i)\Gamma^1$. Thus, the Bloch eigenstates $\ket{u_\mu(\mathbf{k})}$ of the Hamiltonian are also eigenstates of $\mathcal{P}$ at the TRIM. Since $\mathcal{P}^2=\Gamma^0$ and $\mathcal{P}=\mathcal{P}^\dagger$, the eigenvalues associated with $\mathcal{P}$ are $\pm 1$. Therefore, the TRIM points have a definite parity given by the eigenvalues $\xi$ of $\mathcal{P}$. From a simple derivation, shown in App. \ref{ap:final}, one arrives to 
the following expression for the parity of the Kramers degenerate 
occupied bands at the TRIM:
\begin{equation}
    \xi(\mathbf{\Gamma}_i)=-\text{sgn}[d_1(\mathbf{\Gamma}_i)].
    \label{eq: TRIM parity}
\end{equation}
This definition, we find that $\xi=-1$ at $\Gamma$ and at $L_{i=1,2,3}$, 
while being $\xi=1$ at $T$. Furthermore, we find that at the points $L$, $Z$ 
and $Y$ the parity is $\xi=-\text{sgn}[\gamma-1]$ for the distorted 
Hamiltonian. Combining all these results we arrive at the expression 
for the set of indices:
\begin{equation}
\displaystyle
(\nu_0;\nu_1\nu_2\nu_3) = \begin{cases}
(1;111)\ \text{for $\gamma>1$} \\
(0;111)\ \text{for $\gamma<1$}.
\end{cases}
\end{equation}
Hence, if the $[111]$ distorted bond is stronger than the other three, meaning that our system is dimerized, we find that we have a strong topological insulator, meanwhile if the $[111]$ bond is weaker than the other three,
implying that our system is layered, we find that it is a weak 
topological insulator. These results imply that any distortion in the $[111]$ bond transforms our Dirac semimetal into an insulator which can be 
either topologically strong or weak but never topologically trivial.

\section{Effective spin model in the strongly interacting limit}
\label{sec: strongcoupling}
We now analyze our model \eqref{eq: Hubbard model} in the $U \gg \lambda_{SOI},t$ limit. In this case, the model can be mapped onto a Heisenberg-type model:
on a diamond lattice which reads:
\begin{equation}
\mathcal{H} = J \sum_{\langle i,j \rangle } {\bf S}_i \cdot {\bf S}_j +  
J_{SOI} \sum_{\langle \langle i,j \rangle \rangle }(-S^x_i S^x_j -S^y_i S^y_j
+S^z_i S^z_j ),
\label{eq: Heisenberg}
\end{equation}
where $J= {4 t^2 \over U} $ is the standard n.n. AF Heisenberg coupling and $J_{SOI}= {4 \lambda_{SOI}^2 \over U}$ a n.n.n. coupling induced by the spin-orbit coupling. This spin model consists of a FM XY-term and an AFM Z-term which favors antiparallel alignment of the spins similar to the spin model found in Mott insulators in the honeycomb lattices\cite{Rachel2010}.  Note that in the present model we have neglected a possible n.n.n. Heisenberg coupling, $J'= {4 t'^2 \over U}$ since $J'/J << 1$ in the actual (ET)Ag$_4$(CN)$_5$ molecular compound. Indeed, the ground state of the $J-J'$ Heisenberg model on the diamond lattice is N\'eel ordered for $J'/J < {1 \over 8}$ and magnetically disordered for $J'/J > {1 \over 8}$. Hence, we only consider the n.n. $J$ and the n.n.n. $J_{SOI}$ but neglect $J'$. 

Using the Luttinger-Tisza semiclassical approach\cite{Luttinger1946,Luttinger1951,Litvin1974} we analyze the magnetic ground states of model \eqref{eq: Heisenberg}. The main result as discussed in App. \ref{ap:LTA} is  that the N\'eel order which can be pointing in any direction when $J_{SOI}=0$ occurs in the $x-y$ plane when $J_{SOI} \neq 0$. Hence, although N\'eel order is expected in the
presence of SOI, the spins must lie within 
the $x-y$ plane for $\lambda_{SOI} \neq 0$.

\section{Mott transition: slave rotor mean-field theory}
\label{sec: SRMFT}
Since an exact solution of model \eqref{eq: Hubbard model} particularly at intermediate $U/t$ is very challenging approximations are required. Slave-rotor mean-field theory (SRMFT)\cite{Florens2002,Manuel2022,Manuel2024}
allows for a numerically efficient description of the Mott metal-insulator transition. It captures the bandwidth reduction with increasing $U/t$ and is consistent with more sophisticated approaches like Gutzwiller-type wavefunctions, DMFT and variational cluster approaches.\cite{Florens2004}

\subsection{Slave-rotor mean-field theory}
In the slave-rotor approach, the electron creation (annihilation) operator is splitted into a spinless bosonic field carrying only charge (rotor) $e^{i\theta_i}$, and a neutral fermion carrying only spin (spinon), $f^\dagger_{i\alpha}$:
\begin{equation}
c^\dagger_{i\alpha}=f^\dagger_{i\alpha}e^{i\theta_i}\ \ \ (c_{i\alpha}=f_{i\alpha}e^{-i\theta_i}).
\end{equation}

Thereby, model \eqref{eq: Hubbard model} with $M=0$, 
expressed in the slave-rotor formulation reads:
\begin{align}
 \mathcal{H}&=t\sum_{\langle i,j\rangle,\alpha}f_{i\alpha}^\dagger f_{j\alpha} e^{i(\theta_i-\theta_j)}+\frac{U}{2}\sum_iL_i^2
 \nonumber \\
 &+i \lambda_{SOI}\sum_{\langle \langle i,j\rangle\rangle,\alpha,\beta}f_{i\alpha}^\dagger \mathbf{\tau}_{\alpha \beta} \cdot {\mathbf{d}_{il} \times \mathbf{d}_{lj} \over |\mathbf{d}_{il} \times \mathbf{d}_{lj}|} f_{j\beta} e^{i(\theta_i-\theta_j)},
 \nonumber 
\end{align}
where $L_i\ (L_i\equiv n_i-1)$ is the orbital angular momentum that describes the charge quantum number linked to site $i$ and $e^{-i\theta_i}$ acting as a lowering operator of $L_i$. 

Thus, the slave-rotor electron decomposition in spinons and rotors transforms the kinetic energy from a quadratic to a quartic contribution.
Hence, a Hubbard-Stratonovich mean-field approximation is performed to factorize these terms:
\begin{equation}
a_{ij}b_{ij}\sim \langle a_{ij}\rangle b_{ij}+a_{ij}\langle b_{ij}\rangle -\langle a_{ij}\rangle \langle b_{ij}\rangle,
\end{equation}
where we take $a_{ij}=f^\dagger_{i\alpha}f_{j\alpha}$ and $b_{ij}=e^{i(\theta_i-\theta_j)}$. If one further consider the mean-field ansatz, $\ket{\Psi}=\ket{\Psi_f}\ket{\Psi_\theta}$, the hamiltonian can be naturally 
splitted into a spinon $\mathcal{H}_f\equiv\bra{\Psi_{\theta}}\mathcal{H}\ket{\Psi_\theta}$ and a rotor 
$\mathcal{H}_\theta\equiv\bra{\Psi_{f}}\mathcal{H}\ket{\Psi_f}$ hamiltonian, $\mathcal{H}=\mathcal{H}_f+\mathcal{H}_\theta$, with:\cite{thesis}
\begin{align}
\mathcal{H}_f=&t\sum_{\langle i,j\rangle} \chi_{ij}^{f } \sum_{\alpha}f^\dagger_{i\alpha}f_{j\alpha}\nonumber\\
+&i\lambda_{SOI}\sum_{\langle\langle i,j\rangle\rangle}\chi_{ij}^{f}\sum_{\alpha\beta}f_{i\alpha}^\dagger {\mathbf{\tau}_{\alpha \beta} \cdot {\mathbf{d}_{il} \times \mathbf{d}_{lj} \over |\mathbf{d}_{il} \times \mathbf{d}_{lj}|}} f_{j\alpha}\label{eq:spinon_ham1}\\
\mathcal{H}_\theta=&t\sum_{\langle i,j\rangle}\chi_{ij}^{\theta }e^{i(\theta_i-\theta_j)}\notag +\sum_i\frac{U}{2} L_i^2\nonumber\\
+&\lambda_{SOI}\sum_{\langle \langle i,j\rangle\rangle}
\chi_{ij}^{\theta }e^{i(\theta_i-\theta_j)}
\label{eq:theta_ham}
\end{align}
where $\chi_{ij}^{f}\equiv\langle e^{i(\theta_i-\theta_j)}\rangle$ and $\chi_{ij}^{\theta }\equiv\langle \sum_{\alpha}f_{i\alpha}^\dagger f_{j\alpha} \rangle(\langle i\sum_{\alpha\beta}f_{i\alpha}^\dagger {\mathbf{\tau}_{\alpha \beta} \cdot {\mathbf{d}_{il} \times \mathbf{d}_{lj} \over |\mathbf{d}_{il} \times \mathbf{d}_{lj}|}} f_{j\alpha} \rangle)$ for $i,j$ n.n. (n.n.n.).

Under the SRMFT approach, the original model has been mapped onto a free fermion model (renormalized by interactions) $\mathcal{H}_f$ coupled to a quantum $XY$-model for the rotor variables $\mathcal{H}_\theta$. Different approaches with different levels of approximation can be used to solve $\mathcal{H}_\theta$. At the strict local level, since $Z = \chi_{ij}^{f} = 0$, \cite{Florens2002,Florens2004} a trivial paramagnetic Mott insulator consistent with the DMFT prediction survives. Here we take into account short-range spatial electronic correlations by using the soft boson representation \cite{Florens2004} by which $e^{i\theta_i} \rightarrow X_i(\tau)$, imposing the constraint, $|X_i|^2 = 1$, on average. Such SRMFT approach captures intersite electronic correlations present in cluster DMFT \cite{Liebsch2011,Wu2010,Millis2017}
which have been shown to play a crucial role in the Mott transition in semimetals.\cite{Manuel2024} In this approach, the spinons in the Mott insulator, $Z=0$, disperse since $\chi_{ij}^{f} \neq 0$, and form a Fermi surface consisting of the
nodal lines/points in the case of the semimetals considered in the present work\cite{Balents2023}. Hence, our Mott insulator is effectively fractionalized into gapped charge excitations and gapless spinons forming a U(1) Dirac QSL\cite{Fiete2015}.

Based on the approach described above our hamiltonian \eqref{eq: Hubbard model} in the SRMFT approach reads:
\begin{equation}
\mathcal{H}=\mathcal{H}_f+\mathcal{H}_X,
\end{equation}
with:
\begin{align}    
\mathcal{H}_f &=t\sum_{\langle i,j\rangle,\alpha} Q^f_{ij} f^\dagger_{i\alpha}f_{j\alpha} 
\nonumber \\
&+i\lambda_{\text{SOI}}\sum_{\langle\langle i,j\rangle\rangle}\sum_{\alpha\beta}Q^f_{ij}f^\dagger_{i\alpha}\mathbf{\tau}_{\alpha\beta}\cdot\frac{\mathbf{d}_{il}\times \mathbf{d}_{lj}}{|\mathbf{d}_{il}\times \mathbf{d}_{lj}|}f_{j\beta}, \label{eq:Spinon Ham}\\
    \mathcal{H}_X &= t\sum_{\langle i,j\rangle} Q^X_{ij}X_i^*X_j+\sum_i (\frac{U}{2}L^2_i+\rho X_i^*X_i)
  \nonumber \\  
    &+\lambda_{\text{SOI}}\sum_{\langle \langle i,j\rangle\rangle}Q^X_{ij}X_i^*X_j,
    \label{eq:Rotor Ham}
\end{align}
where $\rho$ is the Lagrange multiplier introduced to impose the constraint, $X^*_i X_i=1$, and 
the renormalization factors read:
\begin{eqnarray}
Q^f_{ij} &=& \langle X_i^*X_j\rangle_X,
\nonumber \\
Q^X_{ij} &=& \begin{cases}
\langle \sum_\alpha f^\dagger_{i,\alpha}f_{j\alpha}\rangle_f & \text{$i, j$ are n.n.}  \\
\langle i\sum_{\alpha,\beta}f^\dagger_{i\alpha}\mathbf{\tau}_{\alpha\beta}\cdot\frac{\mathbf{d}_{il}\times \mathbf{d}_{lj}}{|\mathbf{d}_{il}\times \mathbf{d}_{lj}|}f_{j\beta}\rangle_f & \text{$i,j$ are n.n.n.}  
\end{cases}
\nonumber 
\end{eqnarray}
with, $Q^{f,X}_{ij}=Q^{f,X}_t (Q^{f,X}_\lambda)$ for $i,j$ n.n. (n.n.n.), the spinon and rotor renormalization factors. Therefore, the slave-rotor approach allows us to explore the behavior of 
spin-only (spinons) and charge-only (rotors) quasiparticles into which electronic excitations have fractionalized.

From the above, the spinon hamiltonian, $\mathcal{H}_f$, and the kinetic energy contribution to the rotor hamiltonian, $\mathcal{H}_X^{(1)}$ read:
\begin{align}
\mathcal{H}_f(\mathbf{k})&=Q^f_t\sum_{i=1}^2d_r(\mathbf{k})\Gamma^r+Q^f_\lambda\sum_{r=3}^5d_r(\mathbf{k})\Gamma^r,\\
\mathcal{H}_X^{(1)}(\mathbf{k})&=Q^X_t[d_1(\mathbf{k})\sigma^1+d_2(\mathbf{k})\sigma^2]+Q^X_\lambda C(\mathbf{k})\sigma^0,
\end{align}
with:
\begin{align}
    C(\mathbf{k})&= 2i\lambda_{SOI}[\text{cos}(\mathbf{k}\cdot\mathbf{R}_2)+\text{cos}(\mathbf{k}\cdot\mathbf{R}_3) \nonumber\\
    &+\text{cos}(\mathbf{k}\cdot(\mathbf{R}_1-\mathbf{R}_3))+\text{cos}(\mathbf{k}\cdot(\mathbf{R}_2-\mathbf{R}_1))],
\end{align}
where the $\Gamma^r$ are the $\mathcal{PT}$-even Dirac matrices. These expressions reduce to:
\begin{align}    \mathcal{H}_f(\mathbf{k})&=Q_t^f[d_1(\mathbf{k})\sigma^1+d_2(\mathbf{k})\sigma^2]\otimes\tau^0,\label{eq: NLSH}\\
\mathcal{H}_X^{(1)}(\mathbf{k})&=Q_t^X[d_1(\mathbf{k})\sigma^1+d_2(\mathbf{k})\sigma^2],
\end{align}
when $\lambda_{SOI}=0$. Moreover, since our rotor Hamiltonian has 2 bands, a rescaling of $U\rightarrow U/2$ is performed in order to recover the correct atomic limit \cite{Manuel2022}. 

We can now introduce the finite$-T$ Green functions for the spinons and rotors from their corresponding decoupled hamiltonians \eqref{eq:Spinon Ham} and \eqref{eq:Rotor Ham}:  
\begin{align}
 G^{\mu}_f(\mathbf{k},i\omega_n)^{-1}&=i\omega_n-\epsilon_f^{\mu}(\mathbf{k})\label{eq:Green f}\\
 G^\mu_X(\mathbf{k},i\nu_n)^{-1}&=\frac{\nu^2_n}{U}+\rho+\epsilon_X^\mu(\mathbf{k})\label{eq:Green X}
\end{align}
where $\omega_n=(2n+1)\pi/\beta$ and $\nu_n=2n\pi/\beta$, with $n\in \mathbb{Z}$, are the fermionic and bosonic Matsubara frequencies, respectively. $\epsilon_f^{\mu}(\mathbf{k})$ and $\epsilon_X^\mu(\mathbf{k})$ are the dispersion relations of the spinons and rotors and $\mu$ a band index. Note that these dispersion relations contain renormalization effects since they are 
associated with the kinetic energy contributions to $\mathcal{H}_f$ and $\mathcal{H}_X$ which explicitly depend on $Q^f_{ij}$ and $Q^X_{ij}$. 
The derivation of $G_f^{\mu}(\mathbf{k},i\omega_n)$ and  $G_X^{\mu}(\mathbf{k},i\omega_n)$ 
is provided in App. \ref{ap: Green X}.

As shown in App. \ref{ap: self eq}, the renormalization factors, $Q_{ij}^X$ and $Q_{ij}^f$
can be expressed as:
\begin{align}
    Q^f_{ij}&=\frac{1}{N}\sum_{\mu,\mathbf{k}}\eta_i^\mu(\mathbf{k})\eta_j^{\mu*}(\mathbf{k})\frac{U}{2E_X^\mu(\mathbf{k})}[b(E_X^\mu(\mathbf{k}))-b(-E_X^\mu(\mathbf{k}))]\label{eq:Qf sc},\\
    Q^X_{ij}&=\frac{1}{N}\sum_{\mu,\alpha,\mathbf{k}}e^{-i\mathbf{k}\cdot\mathbf{r}_{ij}}\xi_{i\alpha}^{\mu}(\mathbf{k})\xi_{j\alpha}^{\mu*}(\mathbf{k})f(\epsilon_f^{\mu}(\mathbf{k})),\label{eq:Qx sc}
\end{align}
where $\mathbf{r}_{ij}=\mathbf{r}_i-\mathbf{r}_j$ is a vector connecting sites $i$ and $j$ of the lattice, and  $\eta_i^\mu(\mathbf{k})$ and $\xi_{i\alpha}^{\mu}(\mathbf{k})$ are the eigenvectors associated to the kinetic parts of the rotor and spinon Hamiltonians, respectively. Moreover, here $f(x)$ and $b(x)$ represent the Fermi-Dirac and Bose-Einstein distributions and $ E_X^\mu=\sqrt{U(\rho+\epsilon^\mu_X(\mathbf{k}))}$.

Similarly, the constraint can be expressed (see App. \ref{ap: self eq} for details) as:
\begin{align}
    1=\frac{1}{N_cN}\sum_{\alpha,\mathbf{k}}\frac{U}{2E_X^\mu(\mathbf{k})}[b(E_X^\mu(\mathbf{k}))-b(-E_X^\mu(\mathbf{k}))],\label{eq:cons no Z}
\end{align}
where $N_c=2$ is the number of sites in the unit cell.

We have thus obtained a set of three self-consistent equations, \eqref{eq:Qf sc}, \eqref{eq:Qx sc} and \eqref{eq:cons no Z}, from which we compute $Q^f_{ij}$, $Q^X_{ij}$
and $\rho$ for given $U$ and $T$. 

The bosonic nature of the rotors implies the possibility that they can form Bose-Einstein condensates. Within SRMFT approach, the electron quasiparticle weight, $Z$, is directly related with the rotor fraction condensing \cite{Florens2004} at $\mathbf{k}=\Gamma$ (at which the minimum in $\epsilon_X({\bf k})$ occurs). Isolating the $\mathbf{k}=0$-mode in \eqref{eq:Qf sc}  and \eqref{eq:cons no Z}, we write:
\begin{figure*}[t]    
   {\includegraphics[width=0.885\linewidth]{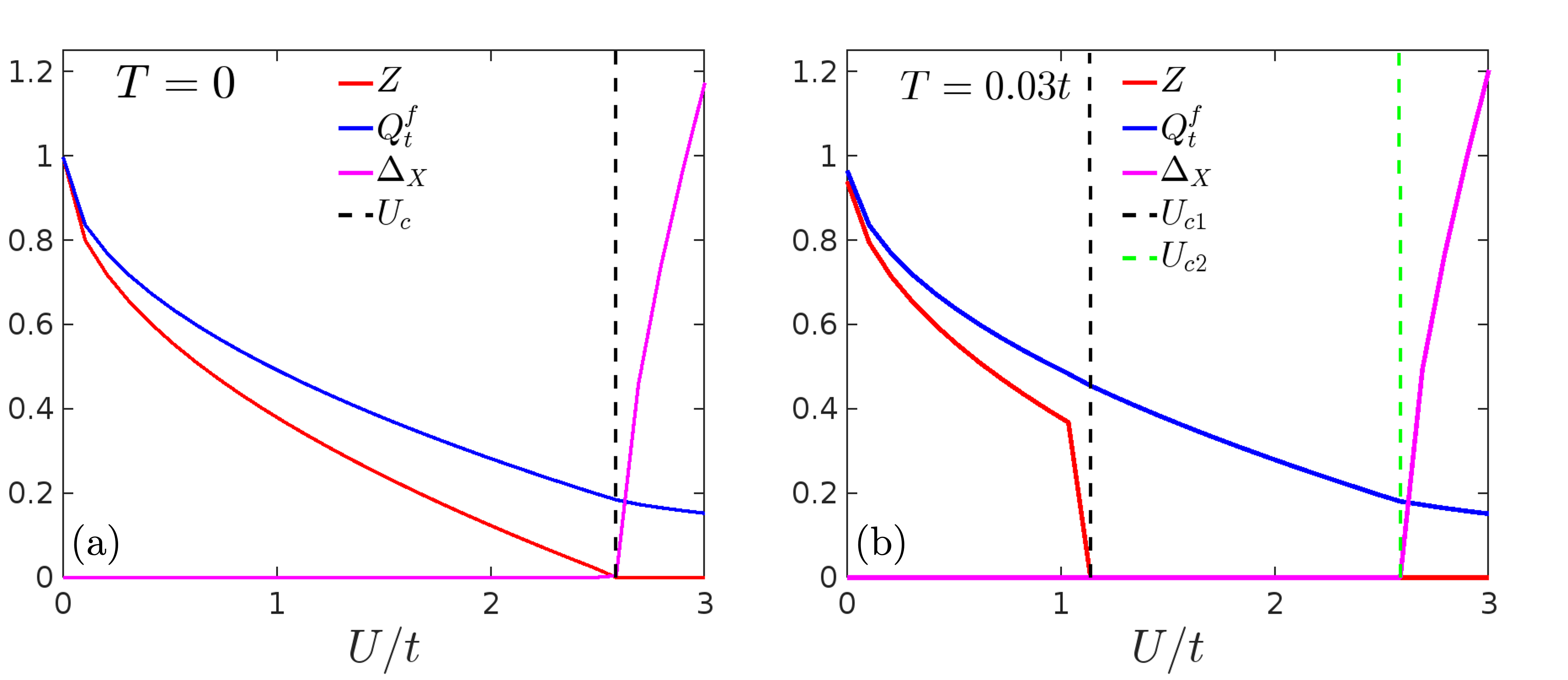}}
    \caption{Mott transition in the Hubbard model \eqref{eq: Hubbard model} on the orthorhombic diamond lattice. The dependence of the quasiparticle weight $Z$, the spinon renormalization factor, $Q^f_t$ and charge gap, $\Delta_X$, on $U$
    are shown at (a) $T=0$ and (b) $T=0.03t$. The Mott insulator is defined by $Z=0$ and $\Delta_X \neq 0$ but $Q_t^f \neq 0$. The vertical dashed lines denote $U_{c}$ at $T=0$ and $(U_{c1}< U_{c2})$ at $T \neq 0$. 
    We have fixed $\gamma=1$ and $\lambda_{SOI}=0$.}
    \label{fig:evolution}  
\end{figure*}

\begin{align}
    Q^f_{ij}=&Z(T)\eta_i^1(0)\eta_j^{1*}(0)
    +\frac{1}{N}\sum_{\mu,\mathbf{k}\neq0}e^{-i\mathbf{k}\cdot\mathbf{r}_{ij}}\eta_i^\mu(\mathbf{k})\eta_j^{\mu*}(\mathbf{k})\notag\\
    &\times\frac{U}{2E_X^\mu(\mathbf{k})}[b(E_X^\mu({\mathbf k}))-b(-E_X^\mu(\mathbf{k}))],\label{eq: Qf}\\
        1=&Z(T)+\frac{1}{N_cN}\sum_{\mu,\mathbf{k}\neq0}\frac{U}{2E_X^\mu({\bf k})}[b(E_X^\mu(\mathbf{k}))-b(-E_X^\mu(\mathbf{k}))],\label{eq: constraint}
\end{align}
where $Z(T)$ has become a new parameter computable from the self-consistent equations.

These equations are iteratively solved as follows. We start with an initial guess for $Q^X_{ij}$, $Q^f_{ij}$ and $\rho$, for which we diagonalize the rotor kinetic Hamiltonian and obtain its eigenvectors $\eta_i^\mu(\mathbf{k})$ and eigenvalues $\epsilon_X^\mu(\mathbf{k})$. From them we recalculate $\rho$ from the constraint equation \eqref{eq: constraint}, before using it to evaluate, in the same equation, the sum over $\mathbf{k}\neq0$ in order to obtain a new value for $Z$. With the knowledge of the new $\rho$ and $Z$ we recalculate the spinon renormalization factor $Q^f_{ij}$, using \eqref{eq: Qf}. We then diagonalize and obtain the eigenvalues $\epsilon_f^{\mu}(\mathbf{k})$ and eigenvectors $\xi_{i\alpha}^{\mu}(\mathbf{k})$ of the spinon kinetic Hamiltonian, which we use for recalculating a new value of $Q^X_{ij}$ through \eqref{eq:Qx sc}. If the recalculated values of $Q^X_{ij}$, $Q^f_{ij}$ and $\rho$ are identical to the initial ones, convergence has been achieved and these are the true self-consistent parameters of our system at a given $U$ and $T$. Otherwise, the process is repeated, reinjecting the recalculated parameters at the beginning of the procedure. The  process is repeated until full convergence is achieved.

At self-consistency, physical properties such as the rotor gap, $\Delta_X \equiv 2\sqrt{U(\rho+\epsilon_X^\mu(\mathbf{k})_{min})}$ can be computed \cite{Florens2004}. A non-zero $\Delta_X$ indicates a bulk charge gap in the systems, ${\it i.e.}$, Mott insulator. 

\subsection{Quantum spin liquid Mott insulator}

Hence, within SRMFT we can encounter two different phases. A semimetallic phase at weak $U$, adiabatically connected to the non-interacting semimetal, characterized by $Z\neq0$ and no charge gap $\Delta_X=0$, and at large-$U$ a nonmagnetic insulating phase characterized by $Z=0$ and non-zero charge gap, $\Delta_X\neq 0$, {\it i.e.} a quantum spin liquid Mott insulator. 

Fig. \ref{fig:evolution} (a) shows the dependence on $U$ of $Z$, $\Delta_X$ and $Q^f_t$
for the Hubbard model on a diamond lattice \eqref{eq: Hubbard model} with no SOI. 
We can see how the quasiparticle weight vanishes, $Z\rightarrow0$, concomitantly with the charge gap opening, $\Delta_X\neq0$, at the critical value of $U_c\sim2.75t$. Therefore, the two mentioned phases can be clearly distinguished here at $T=0$. The nodal loop semimetallic phase ($Z\neq0$ and $\Delta_X=0$) for $U<U_c$ and the Mott insulating phase ($Z=0$ and $\Delta_X\neq0$) for values of $U>U_c$. In the limit $U\rightarrow 0$ 
$Z,Q_f\rightarrow1$, thus recovering the expected non-interacting 
nodal loop semimetal.

In Fig. \ref{fig:evolution} (b) the dependence of the SRMFT parameters at finite temperature, $T=0.03t$, is shown. In contrast $T=0$, at finite-$T$ the transition from the nodal loop semimetal to the Mott insulator becomes a two-step process, with the quasiparticle weight abruptly vanishing at $U_{c1}\sim1.96t$ whereas the charge gap opens up at $U_{c2}\sim2.75t > U_{c1}$. Hence, a new phase is found at $T \neq 0$ between $U_{c1}$ and $U_{c2}$, characterized by $Z=0$ and $\Delta_X=0$ corresponding to a bad smiemetallic phase\cite{Manuel2022,Manuel2024}. 
Interestingly, in this phase $Z=0$ and $Q_t^f\neq 0$, implying that spinons retain the NLSM dispersion of the nearby semimetallic phase. Thus, this phase can be identified as a bad nodal loop semimetal (BNLSM). 

\begin{figure}[h]
    \centering
    \includegraphics[width=0.9\linewidth]{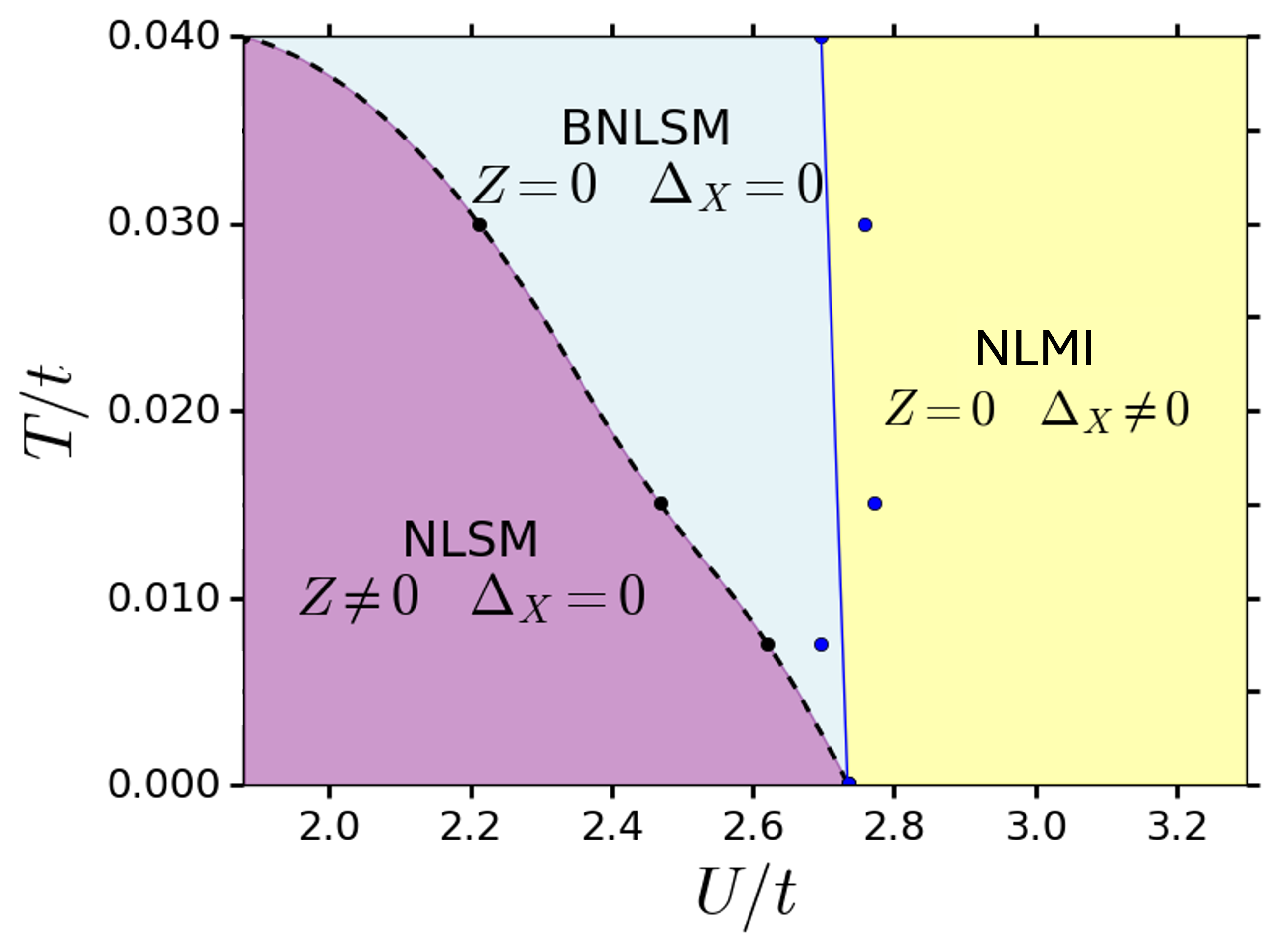}
    \caption{$T-U$ phase diagram of the Hubbard model \eqref{eq: Hubbard model} on an orthorhombic diamond lattice. A bad nodal line semimetal (BNLSM) arises between the NLSM and NLMI phases. The dashed, $U_{c1}(T)$, and solid, $U_{c2}(T)$, lines represent first- and second-order transitions respectively. We have fixed $\gamma=1$ and $\lambda_{SOI}=0$ in this plot.
    \label{fig: U-T first neighbours}}
\end{figure}

The $T-U$ phase diagram obtained from SRMFT is shown in Fig. \ref{fig: U-T first neighbours}. While the rotor gap opens up at $U_{c2}\sim2.7t$ nearly independently of $T$, the quasiparticle weight vanishes, $Z \rightarrow 0$ at a lower $U_{c1}$ which increases with increasing $T$. Hence, the bad semimetallic phase is stabilized in a broader $U/t$ range with increasing $T$. This is consistent with the fact that the sudden drop found in $Z$ at finite-$T$ is due to thermal fluctuations. Since at higher $T$, 
thermal fluctuations are enhanced, the drop in $Z$ occurs sooner.

\subsection{Magnetically ordered Mott insulator}
Since the orthorhombic diamond lattice is bipartite we 
can expect N\'eel type of magnetic order to occur. We explore 
magnetically ordered phases with our slave rotor approach\cite{lee2011,Liu2024} by introducing
a magnetic order parameter $m$ in our model. For N\'eel order we 
introduce the staggered magnetization:
\begin{equation}
    m= {1 \over N}\langle \sum_{i \in A} n_{i\uparrow}- \sum_{i \in B} n_{i\downarrow}\rangle,\label{eq: m}
\end{equation}
 Thus, this parameter describes staggered magnetic order with $S_z$ alternating from positive to negative when going from A to B sites. Thus, $m>0$ indicates N\'eel order sets in the lattice.

Since $m$ involves spin degrees of freedom only, it affects the spinon part of the slave rotor mean-field hamiltonian.\cite{Liu2024} Hence, the hamiltonian is modified as:
\begin{equation}
\mathcal{H} \rightarrow \mathcal{H} + \mathcal{H}^m_f 
\end{equation}
where:
\begin{equation}
    H_{f}^m= -\frac{U}{4}m( \sum_{i \in A} f_{i\uparrow}^\dagger f_{i\uparrow}-\sum_{i \in B} f_{i\downarrow}^\dagger f_{i\downarrow}),\label{eq:m Ham}
\end{equation}
apart from some irrelevant constants. In reciprocal space, this new term reads:
\begin{equation}
    H_f^m(\mathbf{k})=-\frac{U}{4}m\sigma^3\otimes\sigma^3.\label{eq: m Ham rec}
\end{equation}

Taking into account $m$ introduces a new self-consistent equation 
which must be added to our previous original SRMFT equations. 
By transforming \eqref{eq: m} to the reciprocal space and performing the corresponding Matsubara sums, the new equation for $m$ is of the form:
\begin{equation}
    m= \frac{1}{N}\sum_{\mu,\mathbf{k}}[\xi_{i\uparrow}^{\mu*}(\mathbf{k})\xi_{i\uparrow}^\mu(\mathbf{k})-\xi_{i\downarrow}^{\mu*}(\mathbf{k})\xi^\mu_{i\downarrow}(\mathbf{k})]f(\epsilon_f^\mu(\mathbf{k})).
\end{equation}

\begin{figure}[t!]
    \centering
    \includegraphics[width=0.95\linewidth]{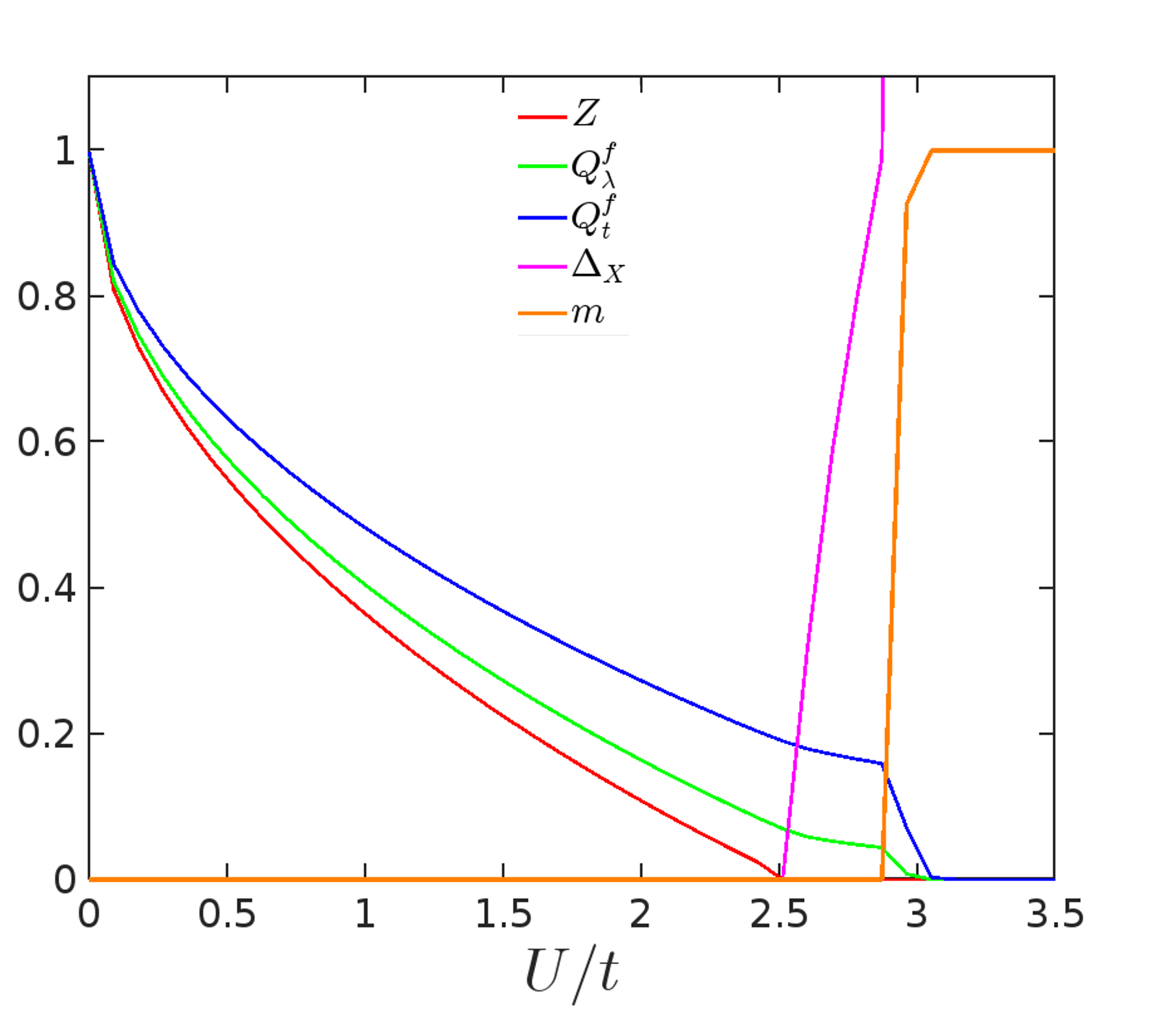}
    \caption{Magnetic order in the Hubbard model on an orthorhombic diamond lattice \eqref{eq: Hubbard model}. The dependence of the N\'eel magnetic moment, $m$, on $U$ is shown together with the quasiparticle weight, $Z$, the renormalization factors,
    $ Q_t^f, Q_\lambda^f$ and the charge gap, $\Delta_X$. Magnetic order, $m \neq 0$, occurs for $U> U_{cm} > U_c$. We have fixed $T=0$, $\gamma=1$ and $\lambda_{SOI}=0.1t$ in this plot. \textcolor{red}{}}
    \label{fig:fig8}
\end{figure}

In Fig. \ref{fig:fig8} we show the dependence of the SRMFT parameters including $m$ with $U$ at $T=0$ for fixed $\lambda_{SOI}=0.2t$. An AFM state emerges 
within the DMI phase {\it i.e.} for $U=U_{cm} > U_{c}$, with the magnetic moment $m$ 
reaching the fully saturated AFM N\'eel ordered state very rapidly. Still, before reaching full saturation, a region with $0<m<1$ and non-zero $Q_t^f, Q_t^\lambda$ arises indicating coexistence of AF and QSL. 

\begin{figure}[t!]
\centering
\includegraphics[width=0.95\linewidth]{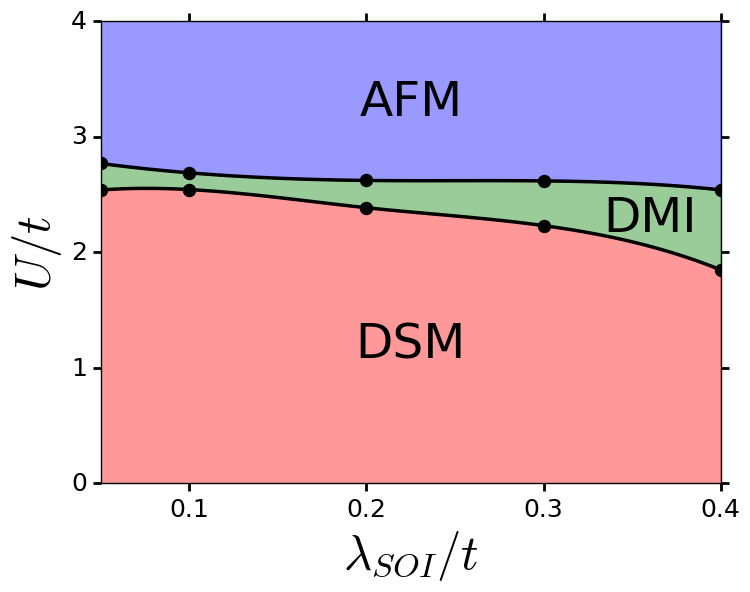}
\caption{$U-\lambda_{SOI}$ phase diagram from slave-rotor mean-field theory including magnetic order. Since AFM N\'eel order arises at $U_{cm}>U_c$ the intermediate DMI phase survives above the DSM. We have fixed $\gamma=1$, $T=0$.}
\label{fig:fig9}
\end{figure}

We have obtained a complete $U$ vs. $\lambda_{SOI}$ phase diagram including AFM order as shown in Fig. \ref{fig:fig9}. Crucially, the DMI spin disordered phase survives in an intermediate $U$ range between the 
DSM and AFM phases in a broad range of $\lambda_{SOI}$ explored.  
As in the phase diagram of Fig. \ref{fig:fig2}, we find that $U_c$ displays a smooth decay with $\lambda_{SOI}$.
This is related to the redistribution of non-interacting 
density of states towards higher energies with $\lambda_{SOI}$ as explained in App. \ref{ap:dependence}.
The (ET)Ag$_4$(CN)$_5$ compounds display a Mott insulating phase with AFM order
at low $T$ and pressures up to around 8 GPa. Above 12 GPa resistivity measurements are consistent with semimetallic behavior. Interestingly, an intermediate phase in the pressure range 8-12 GPa has been interpreted as a disordered Mott insulator \cite{Shimizu2020}. 
From our analysis, we can interpret such intermediate phase observed 
in terms of the DMI found. Further experiments probing magnetism at high pressures are needed to check the nature of the intermediate phase. 
\begin{figure*}[t]
    \centering
    \includegraphics[width=0.85\linewidth]{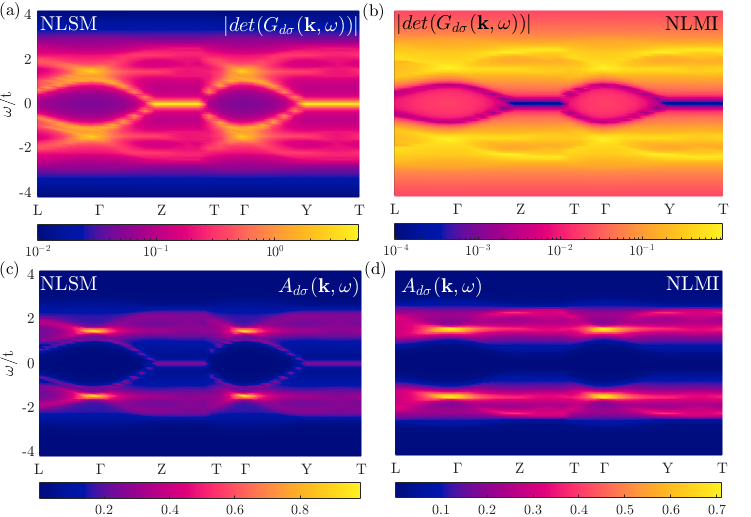}
    \caption{Poles and zeros of the electron Green's function across the Mott transition. The determinant $|det(G_{d\sigma}({\bf k},\omega+i0^+))|$, at $T=0$ and $\lambda_{SOI}=0$ in the (a) NLSM with $U\sim 2t$ and the (b) NLMI with $U\sim 2.9t$. The spectral function, $A_{d\sigma}({\bf k},\omega)$ shown for (c) the NLSM and (d) the NLMI phases. The zeros of the Green's function inside the Mott gap of the NLMI follow the spinon dispersions. }
    \label{fig:Green_spectral}
\end{figure*}

\section{Topology from Green's function zeros}
\label{sec:zeros}
It has been recently shown in an exactly solvable quantum many-body model that the topological properties of a Mott insulator can be obtained from the Green's function zeros \cite{Konig2024}. Interestingly, within a slave-rotor approach to the Kane-Mele Hubbard model the Green's function zeros have been found to follow the underlying spinon dispersion\cite{Sangiovanni2024}. Thus, we characterize the Mott transition on the diamond lattice based on such Green's function zeros perspective. For illustrative purposes, we compare the behavior of the Green's function zeros  with the poles determining the spectra density, $A_{d\sigma}({\bf k} ,\omega)$, across the transition. In other words, we monitor the
topological character of a NLSM as the
Coulomb interaction is increased across the Mott transition solely based on the electronic Green's function.

We first need to obtain the non-local electron Green's function 
from the convolution of the spinon and rotor Green's functions:\cite{Sangiovanni2024, Lee2022, Lee2023} 
\begin{align}
G_{d\sigma}^{ss'}({\bf k },i\omega)&=Z G_{f\sigma}^{ss'}({\bf k },i\omega)\nonumber\\
&+\frac{T}{N}\sum_{{\bf q},i\nu_n}G_{f\sigma}^{ss'}({\bf k-q},i\omega-i\nu_n)G_X^{ss'}({\bf q},i\nu_n),
\label{eq:nonlocalGreen}
\end{align}
where $s,s'=A,B$ and $i\omega=(2n+1)\pi$ ($i\nu_n=2n\pi$) are the fermionic (bosonic) Matsubara frequencies. Note that the Green's function is a $2 \times 2$ matrix. The details on the derivation of the electron Green's functions are given in App. \ref{ap: Green}.

The absolute value of the determinant $| \text{det}G_{d\sigma}({\bf k},\omega+i0^+)|$ is compared with the spectral density, $A_d({\bf k}, \omega)$ along FBZ symmetry directions in Fig. \ref{fig:Green_spectral} both in the NLSM and NLMI phases. Since in the NLSM, $Z \neq 0$, spinons and rotors are combined forming electron quasiparticles leading to poles in the Green's function which dominate the spectra of both $|\text{det}G_{d\sigma}({\bf k},\omega)|$ (Fig. \ref{fig:Green_spectral} (a)) and $A_d({\bf k}, \omega)$ (Fig. \ref{fig:Green_spectral} (c)). Apart from the coherent quasiparticle contribution, one can appreciate the
incoherent Hubbard bands arising from the convoluted rotor-spinon excitations.

In contrast, in the NLMI, strong electronic correlations suppress the quasiparticle weight to zero, $Z=0$, leading to electron fractionalization. The electron fractionalizes into gapless neutral spinons and gapped charged rotors. This is reflected in $|\text{det}G_{d\sigma}({\bf k},\omega)|$ (Fig. \ref{fig:Green_spectral} (b)) and $A_d({\bf k}, \omega)$ (Fig. \ref{fig:Green_spectral} (d)). While the $A_d({\bf k}, \omega)$ and $|\text{det}G_{d\sigma}({\bf k},\omega)|$ spectra are dominated by the Hubbard bands, the zeros of $|\text{det}G_{d\sigma}({\bf k},\omega)|$ inside the Mott gap disperse as the spinon nodal lines characterizing the Fermi surface of the NLSM. Thus, the fractionalized NLMI phase is characterized by the occurrence of gapless Green’s function zeros inside the Mott gap. Since nodal lines characterize the topology of the NLSM, one can automatically associate the topological properties of the NLMI with the zeros of the Green's function. These in turn are found to trace the 
spinon nodal lines obtained within the slave-rotor approach. 

In the NLSM phase, topological drumhead surface states are expected to occur at the projected surface Brillouin zone
following the bulk-surface correspondence. These are analogous to the Fermi arcs arising at the surface of Weyl semimetals. Based on the bulk-surface correspondence established in the strongly interacting regime\cite{Wagner2023,Blason2023}, the NLMI 
should host gapless boundary zeros in the Green's function. Since within slave-rotor theory, the Green's function zeros are directly connected to the gapless boundary spinons, we can conclude that the NLMI 
hosts drumhead surface states of spinons. 
These drumhead surface states differ fundamentally from the metal surface states, which are characterized by poles of the Green’s function. The replacement of poles by zeros illustrates how fractionalization reshapes both bulk and boundary physics, emphasizing the purely spin-driven topology of the NLMI phase. Moreover, intriguing phenomena at NLMI/NLSM interfaces can be expected within slave-rotor theory\cite{Sangiovanni2024}. The gapless boundary spinon modes manifesting as Green’s function zeros would  
cancel the spinon contribution to the NLSM’s surface states 
leaving only drumheads of holons propagating at the NLMI/NLSM interface. 

Our analysis reinforces the relevance of the single particle Green's function, $G_{d\sigma}({\bf k},\omega)$, as a diagnostic tool of the topological properties across the NLSM to NLMI transition\cite{Wagner2023,Konig2024, Sangiovanni2024}. While in the NLSM, conventional quasiparticles are described by the poles as expected, in the NLMI the zeros of $G_{d\sigma}({\bf k},\omega)$ determine the topological properties of the strongly interacting NLMI.

\section{Comparison with observations in (ET)Ag$_4$(CN)$_5$ compounds }
\label{sec: obs}
Experimental observations show how (ET)Ag$_4$(CN)$_5$ compounds 
are Mott insulators up to hydrostatic pressures of about 3 GPa. 
The first important issue is whether a metallic state is induced at 
larger pressures due to the further increase in the bandwidth. Band theory 
predicts that such metallic state is actually a NLSM. Shubnikov-de Haas and 
de Haas-van Alphen oscillations may be used to extract the Fermi surface 
shape and confirm or not their existence\cite{Hussey2020}. The Dirac 
nodal lines could also be probed through ARPES experiments. 
Characteristic drumhead 
surface states extending over the area enclosed by the nodal lines 
arise at the surface Brillouin smoothly connecting to the bulk band crossings. 
Under an applied electric field topological transverse currents 
perpendicular to the applied field associated with opposite Berry phases
on the nodal loop emerge. However, since the contributions 
on opposite sides of the nodal loop cancel there is no net induced current 
unless a filtering device is used\cite{schnyder2018}. 

Based on our SRMFT approach, (ET)Ag$_4$(CN)$_5$ is a Mott insulator that
hosts topological features. Due to spin-charge separation, 
the charge-only excitations are gapped  
while spinons form spinon nodal lines. Since spinons are neutral particles they are difficult 
to detect but thermal instead of charge conductivity experiments 
could be used to probe the nodal lines. On the other hand
spinon drumhead surface states are expected to occur in the Mott insulator within our SRMFT approach. 
They are predicted to lead to a suppression of the bulk Mott gap 
in ARPES experiment probing surface layers onto which the 
nodal loop can be projected. In the present case and based on Fig. \ref{fig:bandstructure}(a), the drumhead surface states will arise on 
the $k_x$-$k_z$  and $k_y$-$-k_z$ surface BZ planes. 
By adsorbing a magnetic impurity in the surface of (ET)Ag$_4$(CN)$_5$
a spinon Kondo-effect can occur in which the magnetic impurity 
forms a singlet with the spinons of the gapless QSLs\cite{Lee2022}. 
This in turn would lead to features in the differential 
conductance measured in STM experiments probing the tunneling
currents through the surface states modified by the presence
of the magnetic impurity. Finally, the magnetic order in our TMI 
would be of the N\'eel 
type always as shown in Fig. \ref{fig: U-T first neighbours}. We 
can expect the N\'eel ordered spins to lay within their $x-y$ plane under SOI.

We now discuss our results in the light of recent experiments on (ET)Ag$_4$(CN)$_5$
at higher pressures reaching 14 GPa \cite{Shimizu2020}. By increasing the 
pressure the electron-electron interaction is effectively reduced. The power 
law dependence of the resistivity $\rho(T) \propto T^{-\alpha}$ detected at low $T$ 
at high pressures with $\alpha$ decreasing from 2.6 to 1.78 as the pressure 
is increased from 8 GPa to 10 GPa. This can indicate the proximity of the 
Mott insulator to a transition to a semimetal. This would be consistent with 
the three possible semimetallic states discussed: a NLSM, a DSM or a Weyl semimetal. However, although SOI is expected to be small in (ET)Ag$_4$(CN)$_5$ crystals it
may be sufficient to open a small gap leading to the topological insulator 
introduced previously. All four weakly interacting phases discussed 
would have specific edge states related to their topology. The NLSM 
would host drumhead surface states, the DSM flat surface states connecting 
the Dirac cones, the WSM Fermi arcs and the topological insulator Z$_2$ surface states which may be distinguished through ARPES and/or STM experiments. 

The topological Mott insulators found in our work are characterized by having spinon surface states that are in a one-to-one correspondence with the surface states 
of the topological electron surface states of the weakly interacting phases. Such spinon surface states could lead to
a closing of the expected bulk gap at certain surfaces of the Mott insulators which could be searched for in ARPES and STM experiments on the TMIs. Alternatively, as recently shown in \cite{Sangiovanni2024}, indirect detection of these spinon topological states can be achieved at the interface between a conventional topological insulator and a TMI. At the NLSM/NLMI interface, the annihilation of the Green’s function zero boundary states with the spin part of the electronic states in 
the semimetal would result in emergent charge-only (holon) drumhead states. In contrast to spinon surface states, these holon drumhead states could be detected in conventional charge transport experiments.

\section{Conclusions and outlook}
\label{sec: conclusions}
We have presented a thorough discussion of the Mott transition in orthorhombic diamond lattices as a platform to access the Mott transition under pressure observed in the molecular compound, (ET)Ag$_4$(CN)$_5$. Our slave rotor analysis predicts a QSL with a charge gap hosting nodal lines (Dirac nodes)  of spinons, the NLMI (DMI) for $\lambda_{SOI}=0$  ($\lambda_{SOI} \neq 0$ ). These Mott insulating phases are topologically non-trivial since they inherit the topological properties of the nearby weakly interacting semimetallic phases 
through the spinon bands. We confirm this picture by obtaining the Green's function zeros which are related to the topological properties of the strongly interacting phases \cite{Wagner2023,Sangiovanni2024}. Since Green's function zeros follow the spinon dispersions, experimental probes of spinons are desirable in order to establish the topological properties of the Mott insulators found here. ARPES experiments should detect 
a suppression of the bulk gap at the surface of these Mott insulators  due to the spinon surface bands \cite{Manuel2024}.
This highlights the crucial role played by Green's function zeros in accessing the topological properties of strongly interacting systems in general.

Magnetic order is explored 
based on an extension of the slave-rotor approach. Although N\'eel order is stabilized in a broad parameter range of the $U$-$\lambda_{SOI}$ phase diagram, a DMI phase survives in an intermediate parameter
regime between the DSM and AFM Mott 
phases. Coexistence of QSL and AFM in the Mott insulator is possible in a rather small parameter range before the magnetic order has reached the fully saturated N\'eel state. 

Our results are broadly consistent with recent observations in (ET)Ag$_4$(CN)$_5$ which indicate a transition from an ambient pressure N\'eel ordered Mott insulator to a semimetallic/semiconducting phase under high pressures. There are indications of an intermediate insulating-like phase which may be interpreted in terms of our DMI phase  arising between the AF Mott insulator and the DSM. 
Depending on SOI, $\lambda_{SOI}$, and degree of dimerization, $\gamma$, different semimetals at high pressures are possible. Even a topological insulator can be favored under sufficient uniaxial pressure along the $[111]$ direction.

Since our work is based on a mean-field approach, future theoretical work should 
consider electronic correlation effects through numerical approaches. An important question is whether the intermediate QSL Mott insulator survives beyond mean-field theory. Future experiments on (ET)Ag$_4$(CN)$_5$ materials should focus on characterizing the intermediate phase arising at pressures around the Mott transition and analyzing the strongly correlated semimetallic phases arising at  larger pressures. The topological aspects of these semimetals could be explored based on magnetic oscillation experiments which can display evidence of non-zero Berry phases associated with the presence of Dirac or line nodes. It is also worth investigating the possibility of inducing superconductivity at even larger pressures.  

\acknowledgments
We acknowledge financial support from (Grant No.
PID2022-139995NB-I00) MICIN/FEDER, Uni\'on Europea,
from the Mar\'ia de Maeztu Programme for Units of Excellence in R\&D (Grant No. CEX2023-001316-M). J. C. acknowledges financial support from the FPI Grant No. PREX2023-000114.

\appendix 
\section{Tight-binding model for (ET)Ag$_4$(CN)$_5$ up to fourth 
n.n. sites}
\label{ap: 4}
For completeness and in order to make our paper self-contained we include the extension of the tight-binding model of the main text up to fourth n. n. sites following previous works\cite{Shimizu2019}.
The orthorhombic diamond lattice structure shown in Fig. \ref{fig:fig1} of (ET)Ag$_4$(CN)$_5$ crystals belong to the non-symmorphic Fddd space and point group: $D_{2h}=\{E,\ C_2(x),\ C_2(y),\ \mathcal{P},\ \sigma(xy),\ \sigma(xz),\ \sigma(yz)\}$, with the center of the lattice being its invariant point. In our work, we have considered the simplest tight-binding model for (ET)Ag$_4$(CN)$_5$ crystals including n.n. hoppings between ET-molecules only. However, in actual
(ET)Ag$_4$(CN)$_5$ crystals further distant hoppings are relevant. 
Table \ref{tab:hoppings} presents hoppings up to the 4$^\text{th}$ n.n. that an element belonging to the sublattice A of the Fddd diamond orthorhombic lattice and located at the origin has. The intensity of the different hoppings is also presented. All the data were collected from \cite{Exp}. Here an x-ray diffraction experiment was conducted to obtain the geometry of the salt while fittings of tight-binding parameters to first-principles calculations where performed to obtain the hoppings intensities. For reproducing the preliminary results shown in that work, we found necessary to exchange the provided values for $t_{2ab}$ and $t_{2ac}$ and change the sign given for $t_{4bc}$. 
\begin{table*}[t]
    \centering
    \caption{List of the nearest neighbors up to $4^{\text{th}}$ order from an A site located at the origin. Order is defined according to the corresponding pair of sublattices, not by the distance, so that the lattice is bipartite. At odd (even) orders two sites belonging to different (same) sublattices are connected with one common distance. The coordinates $(\mathbf{R}_i;\mathbf{d}_{AB})$ are measured from a site A located at the origin. Site B belonging to that same unit cell located at the origin has its corresponding neighbors at $(-\mathbf{R}_i;-\mathbf{d}_{AB})$, as measured from B.  The three lattice parameters of the orthorhombic crystal structure are: $a=13.2150$\AA, $b=19.4783$\AA, and $c=19.6506$\AA.  $\mathbf{R}_1=(0,b/2,c/2)$, $\mathbf{R}_2=(a/2,0,c/2)$, and $\mathbf{R}_3=(a/2,b/2,0)$, are the vectors defining the primitive unit cell of the lattice. } 
    \label{tab:hoppings} 
    \begin{tabular}{cccc}
        \toprule\toprule
        Order\ \ \ & \ \ \ ($\mathbf{R}_i;\mathbf{d}_{AB})$ coordinates\ \ \ &\ \ \ Distance (\AA) \ \ \ &\ \ \  Hopping (meV)\ \ \ \\
        \midrule
1              & $(0,0,0;1)$                  &   $\frac{1}{4}\sqrt{a^2+b^2+c^2}=7.6656$       &    $t_1=-68.442$     \\
1              & $(-1,0,0;1)$                    &   $\parallel$      &    $\parallel$     \\ 
1              &      $(0,-1,0;1)$     &     $\parallel$      &     $\parallel$      \\ 
1              &       $(0,0,-1;1)$              &   $\parallel$        &      $\parallel$     \\ 
2              &           $(0,0,\pm1;0)$            &    $\frac{1}{2}\sqrt{a^2+b^2}=11.7690$       &    $t_{2ab}=-0.487$     \\ 
2              &     $(\pm 1,\mp 1,0;0)$                 &  $\parallel$         &   $\parallel$       \\ 
2              &           $(0,\mp 1,0;0)$            &   $\frac{1}{2}\sqrt{a^2+c^2}=11.8404$       &    $t_{2ac}=4.226$      \\ 
2              &         $(\pm 1,0,\mp 1;0)$              &      $\parallel$     &    $\parallel$      \\ 
3              &           $(-1,1,0;1)$            &   $\frac{1}{4}\sqrt{(3a)^2+b^2+c^2}=12.0863$       &       $t_3=1.966$  \\ 
3              &       $(-1,0,1;1)$                &    $\parallel$       &    $\parallel$     \\ 
3              &         $(1,-1,-1;1)$              &   $\parallel$        &    $\parallel$     \\ 
3              &         $(0,-1,-1;1)$              &  $\parallel$         &    $\parallel$     \\ 
4              &         $(\pm 1,\mp 1,\mp 1;0)$              &    $a=13.2150$      &    $t_{4a}=0.165$     \\ 
4              &           $(\pm 1,0,0;0)$            &   $\frac{1}{2}\sqrt{b^2+c^2}=13.8343$       &    $t_{4bc}=-1.756$     \\ 
4              &          $(0,\pm 1,\mp 1;0)$             &     $\parallel$      &     $\parallel$    \\ 
        \bottomrule\bottomrule
    \end{tabular}
\end{table*}

\begin{figure}[h]
    \centering
    \includegraphics[width=0.8\linewidth]{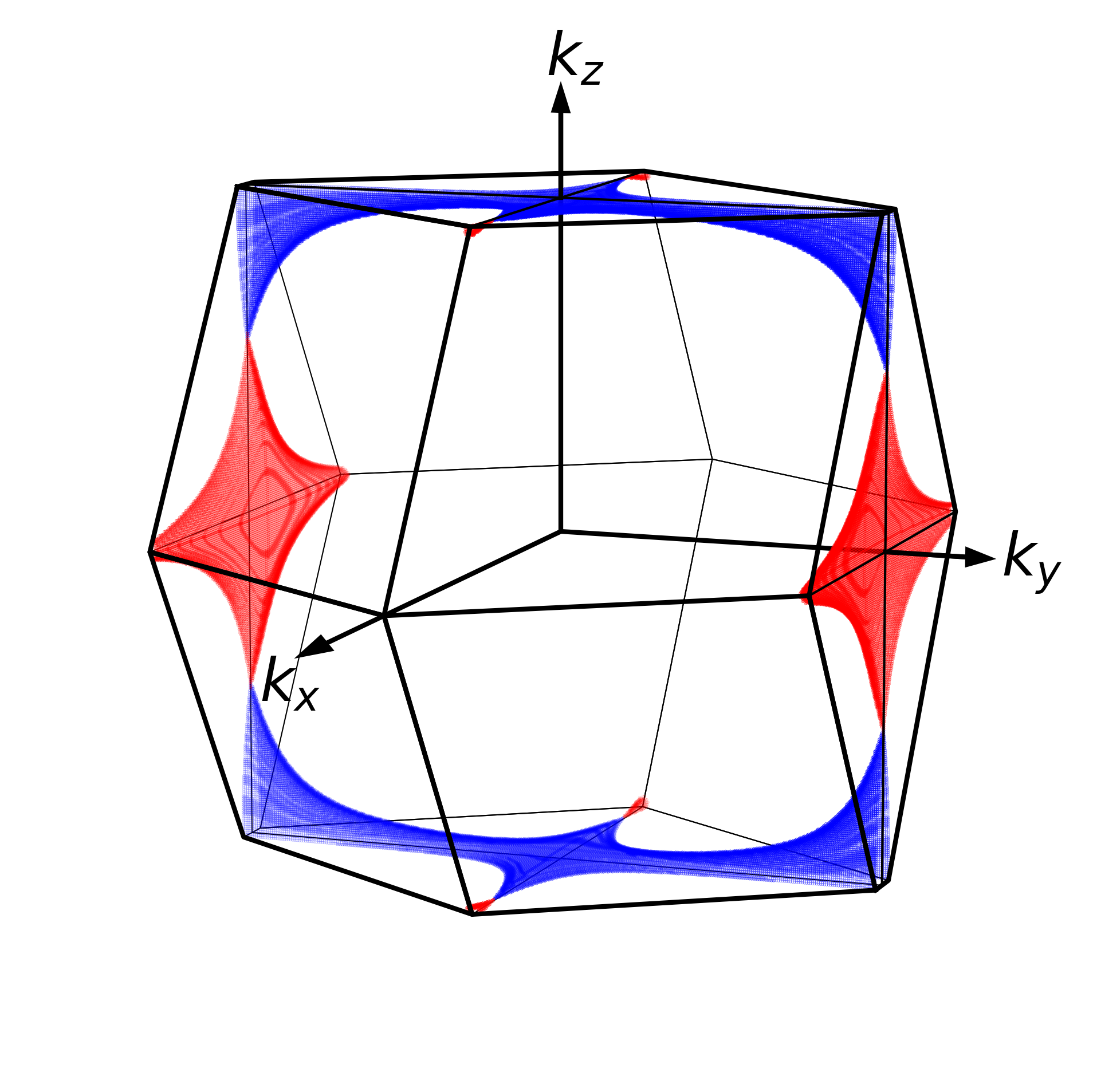}
    \caption{Fermi surface of the actual (ET)Ag$_4$(CN)$_5$ molecular compounds. The Fermi surface for the tight-binding model up to 4th n.n. is shown. In contrast to the n.n. tight-binding model, electron and hole pockets transversing the nodal lines occur.}
    \label{fig:fig11}
\end{figure}

The tight-binding Hamiltonian with real hoppings (no spin dependence) up to 4$^\text{th}$ n.n. reads:
\begin{equation}   
\mathcal{H}_{NLSM}=f_0(\mathbf{k})\sigma^0+f_1(\mathbf{k})\sigma^1+f_2(\mathbf{k})\sigma^2,
\end{equation}
where:
\widetext
\begin{align}
    f_0(\mathbf{k})=&2t_{2ab}[\text{cos}(\mathbf{k}\cdot\mathbf{R}_3)+\text{cos}(\mathbf{k}\cdot(\mathbf{R}_2-\mathbf{R}_1))]+
    2t_{2ac}[\text{cos}(\mathbf{k}\cdot\mathbf{R}_2)+\text{cos}(\mathbf{k}\cdot(\mathbf{R}_3-\mathbf{R}_1))]
    \notag \\&+
    2t_{4a}\text{cos}(\mathbf{k}\cdot(\mathbf{R}_2+\mathbf{R}_3-\mathbf{R}_1))
    +2t_{4bc}[\text{cos}(\mathbf{k}\cdot\mathbf{R}_1)+\text{cos}(\mathbf{k}\cdot(\mathbf{R}_3-\mathbf{R}_2))],
\end{align}
\begin{align}
    f_1(\mathbf{k})=&t_{1}[1+\text{cos}(\mathbf{k}\cdot\mathbf{R}_1)+\text{cos}(\mathbf{k}\cdot\mathbf{R}_2)+(\mathbf{k}\cdot\mathbf{R}_3)]+
    t_{3}[\text{cos}(\mathbf{k}\cdot(\mathbf{R}_1-\mathbf{R}_2))+\text{cos}(\mathbf{k}\cdot(\mathbf{R}_1-\mathbf{R}_3))\notag \\ & +\text{cos}(\mathbf{k}\cdot(\mathbf{R}_2+\mathbf{R}_3))+\text{cos}(\mathbf{k}\cdot(\mathbf{R}_3+\mathbf{R}_2-\mathbf{R}_1))],
\end{align}
\begin{align}
    f_2(\mathbf{k})=&t_{1}[\text{sin}(\mathbf{k}\cdot\mathbf{R}_1)+\text{sin}(\mathbf{k}\cdot\mathbf{R}_2)+(\mathbf{k}\cdot\mathbf{R}_3)]+
    t_{3}[\text{sin}(\mathbf{k}\cdot(\mathbf{R}_1-\mathbf{R}_2))+\text{sin}(\mathbf{k}\cdot(\mathbf{R}_1-\mathbf{R}_3))\notag \\ & +\text{sin}(\mathbf{k}\cdot(\mathbf{R}_2+\mathbf{R}_3))+\text{sin}(\mathbf{k}\cdot(\mathbf{R}_3+\mathbf{R}_2-\mathbf{R}_1))].
\end{align}
\twocolumngrid 
Recall that $\sigma^0$ is the identity matrix and $\sigma^{i=1,2}$ are the Pauli matrices.
In contrast to the n.n. tight-binding model in which the Fermi surface consists of Dirac nodal lines, the Fermi surface of the tight-binding model up to fourth n.n. sites consists of hole and electron pockets \cite{Shimizu2019} as shown in Fig. \ref{fig:fig11}. 
Note that these pockets enclose nodal lines with non-zero Berry phases, $\zeta_1 = \pm \pi $, which although may not lead to observable features in magnetic oscillatory phenomena due to cancellation of Berry phases around electron or hole extremal orbits, it can nevertheless lead to effects in the Landau level spectra \cite{Moessner2018}.
\section{Dirac semimetal model.}\label{ap: DSM}

The consideration of a spin dependency with the introduction of the Fu-Kane-Mele spin orbit interaction in \eqref{eq:DSM} leads to the need of considering the four-dimensional tensor product space $\mathfrak{L}\otimes\mathfrak{G}$ for correctly describing the system. A basis of this space is given by the Kronecker product of the sublattice and spin basis: $\{A,B\}\otimes\{\uparrow,\downarrow\}=\{A\uparrow,A\downarrow,B\uparrow,B\downarrow\}$. This ordering arranges the four $2\times2$ connected blocks of the matrix representations in this basis, with each corresponding to a fixed pair of sublattices. We refer to these as spin blocks, as the spin degrees of freedom vary across them. 

When Fourier transforming \eqref{eq:Dirac H} to the reciprocal space, its matrix representation in the previous basis becomes, in $2\times2$ spin blocks:
\[\displaystyle
\mathcal{H}_{DSM}(\mathbf{k}) = \left(
\begin{array}{c|c}
\begin{array}{cc}
\mathcal{H}_{AA}(\mathbf{k}) \\

\end{array} &
\begin{array}{cc}
\mathcal{H}_{AB}(\mathbf{k})  \\

\end{array} \\
\hline
\begin{array}{cc}
\mathcal{H}_{BA}(\mathbf{k})\\

\end{array} &
\begin{array}{cc}
\mathcal{H}_{BB}(\mathbf{k})  \\

\end{array}
\end{array}
\right)
\]
where the different matrix blocks read:
\begin{equation}
    \mathcal{H}_{AB}(\mathbf{k})=\mathcal{H}_{BA}^\dagger(\mathbf{k})=(d_1(\mathbf{k})+i d_2(\mathbf{k}))\sigma^0, \notag
\end{equation}
\begin{align}
&\mathcal({H}_{AA})_{11}(\mathbf{k})=-\mathcal({H}_{AA})_{22}(\mathbf{k})=-\mathcal({H}_{BB})_{11}(\mathbf{k})=\mathcal({H}_{BB})_{22}(\mathbf{k}) 
\nonumber \\
&=\frac{2a\lambda_{SOI}}{\sqrt{a^2+c^2}}\left[\text{sin}(\mathbf{k}\cdot\mathbf{R}_2)+\text{sin}(\mathbf{k}\cdot(\mathbf{R}_1-\mathbf{R}_3))\right], \notag
\end{align}

\begin{align}
&\mathcal({H}_{AA})_{12}(\mathbf{k})=\mathcal({H}_{AA})_{21}^*(\mathbf{k})=-\mathcal({H}_{BB})_{12}(\mathbf{k})=-\mathcal({H}_{BB})_{21}^*(\mathbf{k})
\notag \\
&=\frac{-2c\lambda_{SOI}}{\sqrt{a^2+c^2}}\left[\text{sin}(\mathbf{k}\cdot\mathbf{R}_2)-\text{sin}(\mathbf{k}\cdot(\mathbf{R}_1-\mathbf{R}_3))\right]
\nonumber \\
&+\frac{2\lambda_{SOI}}{\sqrt{a^2+b^2}}[-(b-ia)[\text{sin} (\mathbf{k}\cdot(\mathbf{R}_1-\mathbf{R}_2))+(b+i a) \text{sin}(\mathbf{k}\cdot\mathbf{R}_3)]. \notag
\end{align}

Here $i$ refers to the imaginary unit. Notice that the notation ${1,2}\equiv {\uparrow,\downarrow}$ is followed and that the functions $d_1(\mathbf{k})$ and $d_2(\mathbf{k})$ are once again \eqref{eq: d1 NLSM hamiltonian} and \eqref{eq: d2 NLSM hamiltonian}, respectively. As explained in the text, this Hamiltonian can be rewritten in the compact form:
\begin{equation}
    \mathcal{H}_{DSM}(\mathbf{k})=\sum_{i=1}^5d_i(\mathbf{k})\Gamma^i,
\end{equation}
where $\Gamma^i$ are the $\mathcal{PT}$-even Dirac matrices introduced in \eqref{eq: Dirac matrices}. From the definition of this matrices and by simply comparing with the $4\times 4$ Hamiltonian previously introduced, the different coefficient functions $d_i(\mathbf{k})$, that still remain unknown to us, can be shown to be:
\widetext
\begin{align}
d_3(\mathbf{k})=\frac{-2c\lambda_{SOI}}{\sqrt{a^2+c^2}}\left[\text{sin}(\mathbf{k}\cdot\mathbf{R}_2)-\text{sin}(\mathbf{k}\cdot(\mathbf{R}_1-\mathbf{R}_3))\right]&+\frac{2b\lambda_{SOI}}{\sqrt{a^2+b^2}}\left[\text{sin}(\mathbf{k}\cdot\mathbf{R}_3)-\text{sin}(\mathbf{k}\cdot(\mathbf{R}_1-\mathbf{R}_2))\right], \notag \\
d_4(\mathbf{k})= \frac{2a\lambda_{SOI}}{\sqrt{a^2+b^2}}[\text{sin}(\mathbf{k}\cdot\mathbf{R}_3)&+\text{sin}(\mathbf{k}\cdot(\mathbf{R}_1-\mathbf{R}_2)) ],   \notag \\
    d_5(\mathbf{k})= \frac{2a\lambda_{SOI}}{\sqrt{a^2+c^2}}[\text{sin}(\mathbf{k}\cdot\mathbf{R}_2)&+\text{sin}(\mathbf{k}\cdot(\mathbf{R}_1-\mathbf{R}_3))].  
\end{align}

\twocolumngrid

As shown in \cite{PhysRevB.97.161113}, a null value of $\zeta_1$  will be an indicator of the degeneracy being purely accidental and removable by any small perturbation in the Hamiltonian preserving all its symmetries. On the other hand, from a non-zero value of the Berry phase we can infer that the nodal loop is protected by the SU(2) and $\mathcal{PT}$ symmetries of the system. 

Therefore, the easiest way to prove if this first topological index is zero is by slightly perturbing our Hamiltonian while preserving all its symmetries, and then study the persistence of the nodal loops. For doing so, we rewrite the function $d_1(\mathbf{k})$ given in \eqref{eq: d1 NLSM hamiltonian}, which defines our Hamiltonian, as:
\begin{equation}
d_1(\mathbf{k})=t\left[\gamma+\text{cos}(\mathbf{k}\cdot\mathbf{R}_1)+\text{cos}(\mathbf{k}\cdot\mathbf{R}_2)+\text{cos}(\mathbf{k}\cdot\mathbf{R}_3)\right],
\end{equation}
where $\gamma\in \mathbb{R}$ now distorts the hopping between elements belonging to the same unit cell in the $[111]$ direction, relative to the hoppings between different unit cells. This leads to bond dimerization along the $[111]$ direction of the lattice.

\section{Parity at the TRIM in the Dirac Semimetal}\label{ap:final}

We present here the derivation of the parity associated to the pair of occupied Kramers degenerate bands provided by the Dirac semimetallic Hamiltonian (Fu-Kane-Mele model) \eqref{eq:DSM} at a TRIM $\mathbf{\Gamma}_i$. As discussed in the text, this Hamiltonian reduces at $\mathbf{\Gamma}_i$ to  
\begin{equation}
    \mathcal{H}_{DSM}(\mathbf{\Gamma}_i)=d_1(\mathbf{\Gamma}_i)\Gamma^1.\notag
\end{equation}

Therefore, considering $\ket{u_-(\mathbf{k})}$ as the eigenstate that describes the pair of Kramer degenerate occupied bands with an associated eigenergy $E_-(\mathbf{k})$,

\begin{equation}
\mathcal{H}_{DSM}(\mathbf{\Gamma}_i)\ket{u_-(\mathbf{\Gamma}_i)}=E_-(\mathbf{\Gamma_i})\ket{u_-(\mathbf{\Gamma}_i)}
\end{equation}
which implies:
\begin{equation}
\Rightarrow \Gamma^1\ket{u_-(\mathbf{\Gamma}_i)}=\frac{|d_1(\mathbf{\Gamma}_i)|}{-d_1(\mathbf{\Gamma}_i)}\ket{u_-(\mathbf{\Gamma}_i)},\label{eq:derivation}
\end{equation}

where an explicit expression of $E_-(\mathbf{k})$ \eqref{eq: Dirac energy} has been taken into consideration. Recalling that in this model the parity at a TRIM is definite and determined by the eigenvalues $\xi(\mathbf{\Gamma}_i)$ of $\mathcal{P}$, and that $\mathcal{P}=\Gamma^1=\sigma^1\otimes\tau^0$, it is easy to see from \eqref{eq:derivation} that

\begin{equation}
\xi(\mathbf{\Gamma}_i)=-\text{sgn}[d_1(\mathbf{\Gamma}_i)],
\end{equation}

which is the expression provided in the text \eqref{eq: TRIM parity} for the parity of the Kramers degenerate occupied bands associated to \eqref{eq:DSM} at a TRIM.

\section{Luttinger-Tisza approximation}
\label{ap:LTA}

Here we provide details on the Luttinger-Tisza approximation used to analyze the magnetic properties our Heisenberg-type model 
(\ref{eq: Heisenberg}) on the diamond lattice. First, spin operators are Fourier transformed: 
\begin{equation}
S^\alpha_{i,s} = {1 \over \sqrt{N}} \sum_{\bf q} e^{i {\bf k} \cdot {\bf R}_i } S^\alpha_{{\bf k}, s},
\end{equation}
where ${\bf R}_i$ denotes the position of the unit cells on the diamond lattice (see Fig. \ref{fig:fig1} and Table \ref{tab:hoppings}), $s$ denotes the sublattice type and $N$ the 
number of sites on each sublattice. We assume that there are only two sublattices, 
$s=A, B$, as in the diamond lattice considered here.

Our model \eqref{eq: Heisenberg} in ${\bf k}$ space reads:
\begin{equation}
H= \sum_{{\bf k}, \gamma, s, s'} S^\alpha_{{\bf k}, s} \Lambda_{ss'}^\alpha({\bf k}) S^\alpha_{-{\bf k}, s'},
\end{equation}
where the three $2 \times 2$, $\Lambda^\alpha({\bf k})$, with the $\alpha=x,y, z$ matrices are expressed as:
\begin{equation}
\Lambda^\alpha({\bf k})=
\begin{pmatrix}
\Lambda^\alpha_{AA}({\bf k}) & \Lambda^\alpha_{AB}({\bf k}) \\
 \Lambda^{\alpha*}_{AB}({\bf k}) & \Lambda^\alpha_{BB}({\bf k}) \\
\end{pmatrix}
\end{equation}
 with:
 \begin{eqnarray}
 \Lambda^\alpha_{AB}({\bf k}) &=&\frac{J}{2} \left[  e^{-i {\bf k} \cdot {\bf R}_1} +  
 e^{-i {\bf k} \cdot {\bf R}_2} +  e^{-i {\bf k} \cdot {\bf R}_3}  + 1\right],
 \nonumber \\ 
 \Lambda^{x/y}_{ss}({\bf k}) &=& -J_{SOI} [ \cos({\bf k} \cdot {\bf R}_3) + \cos( {\bf k} \cdot ({\bf R}_2-{\bf R}_1)) 
 \nonumber \\
 &+& \cos({\bf k} \cdot {\bf R}_2) + \cos({\bf k} \cdot ({\bf R}_3-{\bf R}_1)) ],
 \nonumber \\
 \Lambda^z_{ss}({\bf k}) &=& -\Lambda^{x/y}_{ss}({\bf k}),
 \end{eqnarray}
 with ${\bf R}_1=(0,{b \over 2}, {c \over 2})$, ${\bf R}_2=( {a \over 2}, 0, {c \over 2} )$,  ${\bf R}_3=({a \over 2}, {b \over 2},0 )$.
 The Luttinger-Tisza condition on the absolute spin magnitude 
 of the whole lattice reads:
 \begin{equation}
 \sum_{{\bf k},n} {\bf S}_{{\bf k},n} 
 \cdot {\bf S}_{-{\bf k},n} = N_c N S^2,   
 \end{equation}
The constraint is introduced through a single Lagrange 
multiplier, $\lambda$, in the free energy: 
$F=H-\lambda (\sum_{{\bf k},n} {\bf S}_{{\bf k},n}\cdot{\bf S}_{-{\bf k},n} - N_c N S^2 )$. The minimization of $F$ leads to a set of self-consistent equations:
 \begin{equation}
 \sum_{m} \Lambda^\alpha_{nm}({\bf k}) S^\alpha_{{\bf k}, m} = \lambda S_{{\bf k}, n}^\alpha. 
 \end{equation}
 Hence, from the diagonalization of each $\Lambda^\alpha({\bf k})$ matrix, we obtain a 
 set of eigenvalues $\lambda$. For a given eigenvalue, the energy of the system can be expressed as:
 \begin{eqnarray}
 H&=&\sum_{{\bf k}, \alpha, n}  \left( \sum_m  \Lambda^\alpha_{nm}({\bf k}) S^\alpha_{{\bf k}, m}  \right)  S^\alpha_{-{\bf k},n} 
 \nonumber \\
 &=& \lambda \sum_{{\bf k}, n} S^\alpha_{{\bf k}, n} S^\alpha_{-{\bf k}, n}=  \lambda N_c N S^2.
 \end{eqnarray}

So the energy per unit cell of the system is given by the lowest
 $\lambda$ common to all three $\Gamma^\alpha$ matrices. 
 The ground state energy is given by the lowest  $\lambda$ on 
the 1st Brillouin zone.

We discuss the two relevant cases:

\subsection{$J_{SOI}=0$}

\begin{figure}[h]
    \centering
    \includegraphics[width=0.9\linewidth]{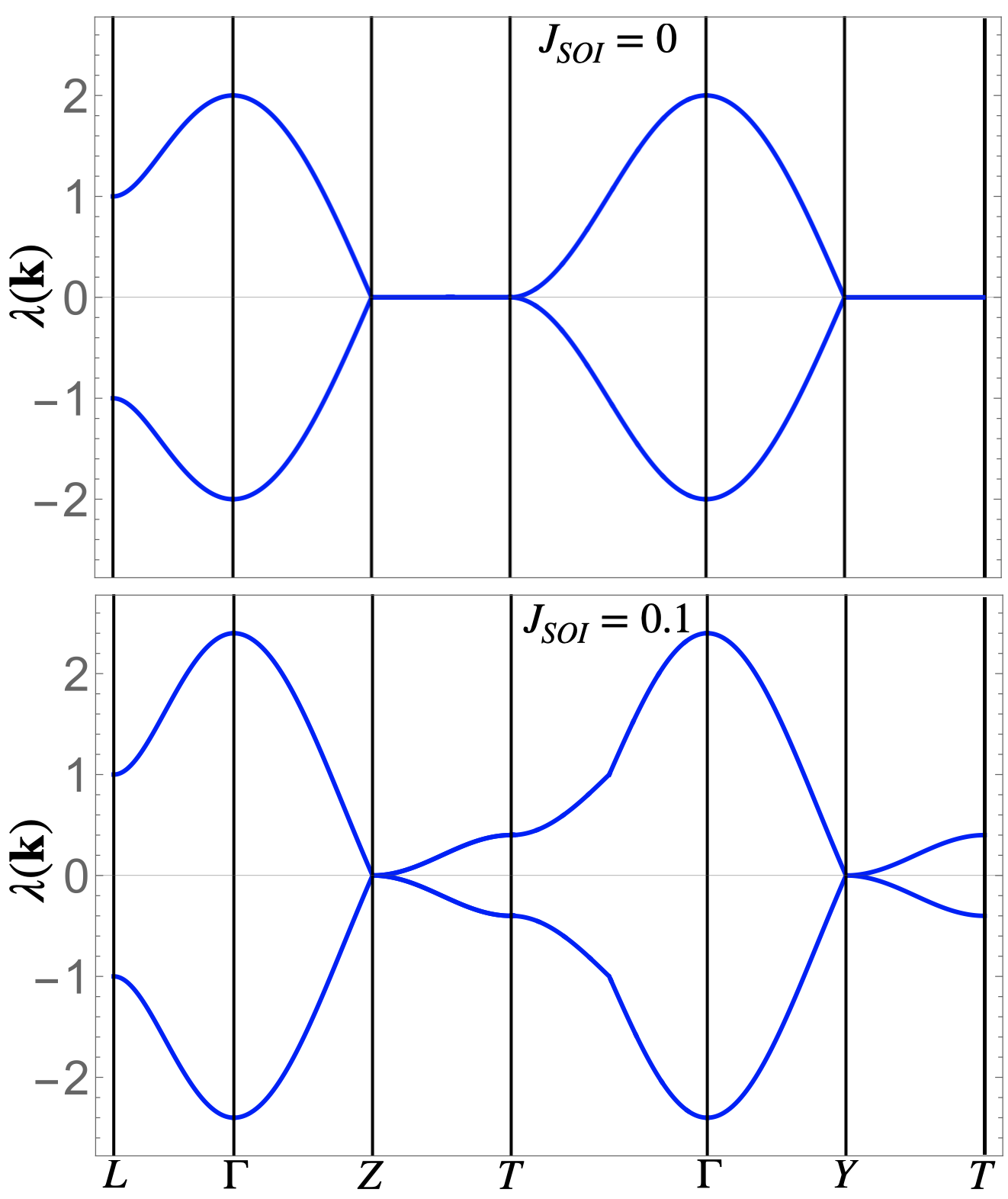}
    \caption{Eigenvalues obtained from the Luttinger-Tisza approach on the Kitaev-Heisenberg model. The ${\bf k}$ dependence of $\lambda(k)$ eigenvalues are shown for the spin model (\ref{eq: Heisenberg}) in the main text. We take $J=1$ in these plots }
    \label{fig:LTA}
\end{figure}

As can be observed in Fig. \ref{fig:LTA} the lowest eigenvalue of the 
$\Lambda^\alpha( {\bf k})$ matrix, $\lambda_{-}({\bf k})$ attains its minimum value at the $\Gamma$-point, $\lambda_{-}(\Gamma)=-2 J$.
Hence, the ground state of the system reads:
\begin{equation}
{E_0 \over N}= 2 S^2 \lambda_-(\Gamma).
\end{equation}

In this case the ground state eigenvector is:
\begin{equation}
S^\alpha_{{\bf Q}=\Gamma,s} = (-1)^s
\end{equation}
with $s=0$ for A sites and $s=1$ for B sites and $\alpha=x,y,z$

\subsection{$J_{SOI} \neq 0$ }

While the minimum still occurs at the $\Gamma$-point we now
have that the eigenvector is different:
\begin{eqnarray}
S^x_{{\bf Q}=\Gamma,s} = S^y_{{\bf Q}=\Gamma,s}= (-1)^s,
\nonumber \\
S^z_{{\bf Q}=\Gamma,s}=0.
\end{eqnarray}
This means that although N\'eel order persists for 
$J_{SOI} \neq 0$ the spins lie within the $x-y$ plane. 
This is in contrast to the $J_{SOI}=0$ case for which 
due to the SU(2) symmetry, the N\'eel order can point in 
any direction. 

\section{Rotor Green Function}\label{ap: Green X}

As commented in the text, obtaining $G_f^{\mu}(\mathbf{k},i\omega_n)$ from the spinon Hamiltonian \eqref{eq:Spinon Ham} is straightforward while the evaluation of $G_X^\mu(\mathbf{k},i\nu_n)$ from the rotor Hamiltonian \eqref{eq:Rotor Ham} requires a bit more of work. First, we need to recall that the rotor Hamiltonian reads:
\begin{equation}
    \mathcal{H}_X=\sum_i(\frac{U}{2}L_i^2+\rho X_i^*X_i)+t\sum_{\langle i,j\rangle}Q_{ij}^XX_i^*X_j=\mathcal{H}_X^{(0)}+\mathcal{H}_X^{(1)}.
\end{equation}
Here $\mathcal{H}^{(0)}$ is the local contribution of the Hamiltonian, which can be identified as the strong-coupling Bose-Hubbard Hamiltonian, whereas $\mathcal{H}^{(1)}$ is the interaction part of the full Hamiltonian. Considering that $L_i^2$ can be written as $L_i^2=\partial^2_\tau X_i^*X_i$, with $\tau$ being the imaginary time associated to the interaction representation, one can easily see that the inverse of the zero order Green's function reads: 
\begin{equation}
    G_X^{(0)}(\mathbf{k},i\nu_n)^{-1}=\frac{\nu_n^2}{U}+\rho,
\end{equation}
since the Fourier transform of $\partial^2_\tau$ is $\nu^2$. This new representation of $L_i^2$ is an important result that follows from the treatment of the Hubbard model under the SRMFT approach within the Lagrangian formalism \cite{Florens2002}. On the other hand, the self-energy associated to $\mathcal{H}^{(1)}$ is the kinetic energy dispersion of the diamond orthorhombic lattice with the hoppings renormalized by $Q^X_{ij}$, $\Sigma^\mu_1=\epsilon_X^\mu$, where $\mu$ refers once again to the band index. Therefore, making use of the Dyson equation:
\begin{align}
    G_X^\mu(\mathbf{k})^{-1} &=G^{(0)}(\mathbf{k},i\nu_n)^{-1}-G^{(0)}(\mathbf{k},i\nu_n)\Sigma_1^\mu G^{(0)}(\mathbf{k},i\nu_n)^{-1}
    \nonumber \\
    &=\frac{\nu_n^2}{U}+\rho+\epsilon_X^\mu(\mathbf{k}),
\end{align}
the rotor Green function introduced in the text  \eqref{eq:Green X} is recovered. It is important to note that expanding our Hubbard Hamiltonian by incorporating terms such as a spin-orbit interaction will result in the same rotor and spinon Green functions, as the effects of these additional terms will only impact the eigenvalues $\epsilon_X^\mu(\mathbf{k})$ and $\epsilon_f^{\mu}(\mathbf{k})$ associated to the kinetic parts of the rotor and spinon Hamiltonians respectively. 

\section{Self-consistent equations}\label{ap: self eq}

The renormalization factors that characterize the spinon \eqref{eq:Spinon Ham} and rotor \eqref{eq:Rotor Ham}  Hamiltonians provided in the text are $Q^f_{ij}=\langle X_i^*X_j\rangle_X$ and $Q^X_{ij}=\langle\sum_\alpha f^\dagger_{i\alpha}f_{j\alpha}\rangle_f$ respectively. Expressing them in the reciprocal space, one finds:
\widetext
\begin{align}
    Q^f_{ij}&=\frac{1}{N}\sum_{\mathbf{k}}e^{i\mathbf{k}\cdot\mathbf{r}_{ij}}\langle X^*_i(\mathbf{k}) X_j(\mathbf{k})\rangle_f
    =\frac{1}{N}\sum_{\mu,\mathbf{k}}e^{-i\mathbf{k}\cdot\mathbf{r}_{ij}}\eta_{i}^\mu(\mathbf{k})\eta_{j}^{\mu*}(\mathbf{k})\langle X_\mu^*(\mathbf{k})X_\mu(\mathbf{k})\rangle_f \notag \\
    &=\frac{1}{N}\sum_n\sum_{\mu,\mathbf{k}}e^{-i\mathbf{k}\cdot\mathbf{r}_{ij}}\eta_i^\mu(\mathbf{k})\eta_j^{\mu*}(\mathbf{k})\frac{1}{\nu^2_n/U+\rho+\epsilon_X^\mu(\mathbf{k})},\label{eq: ap Qf}
\end{align}

\begin{align}
        Q^X_{ij}&=\frac{1}{N}\sum_{\alpha,\mathbf{k}}e^{-i\mathbf{k}\cdot\mathbf{r}_{ij}}\langle f_{i\alpha}^\dagger(\mathbf{k})f_{j\alpha}(\mathbf{k})\rangle_X 
    =\frac{1}{N}\sum_{\mu,\alpha,\mathbf{k}}e^{-i\mathbf{k}\cdot\mathbf{r}_{ij}}\xi_{i\alpha}^{\mu}(\mathbf{k})\xi_{j\alpha}^{\mu*}(\mathbf{k})\langle f^\dagger_{\mu}(\mathbf{k})f_{\mu}(\mathbf{k})\rangle_X\notag \\
    &=\frac{1}{N}\sum_n\sum_{\mu,\alpha,\mathbf{k}}e^{-i\mathbf{k}\cdot\mathbf{r}_{ij}}\xi_{i\alpha}^{\mu}(\mathbf{k})\xi_{j\alpha}^{\mu}(\mathbf{k})\frac{1}{i\omega_n-\epsilon^{\mu}_f(\mathbf{k})},\label{eq:Qx A}
\end{align}
\twocolumngrid 
where $\mathbf{r}_{ij}=\mathbf{r}_i-\mathbf{r}_j$ is a vector connecting sites $i$ and $j$ of the lattice, and  $\eta_i^\mu(\mathbf{k})$ and $\xi_{i\alpha}^{\mu}(\mathbf{k})$ are the eigenvectors associated to the kinetic parts of the rotor and spinon Hamiltonians respectively. Notice that the sums over the corresponding Matsubara frequencies have been performed:
\begin{align}
    \langle f^\dagger_{\mu} f_{\mu}\rangle_f=\frac{1}{\beta}\sum_nG_f^\mu(\mathbf{k},i\omega_n),\\
    \langle X_\mu^*(\mathbf{k})X_\mu(\mathbf{k})\rangle_X=\frac{1}{\beta}\sum_nG_X^\mu(\mathbf{k},i\nu_n).
\end{align}
Realizing that the rotors Green function \eqref{eq:Green X} can be written as the propagator in the quantum harmonic oscillators with energies $\pm E_X^\mu(\mathbf{k})=\pm\sqrt{U(\rho+\epsilon_X^\mu(\mathbf{k}))},$
\begin{align}
   & G_X^\mu(\mathbf{k},i\nu_n)=\frac{1}{\nu^2/U+\rho+\epsilon_X^\mu(\mathbf{k})}
    \nonumber \\
    &=\frac{U}{2E_X^\mu(\mathbf{k})}\left(\frac{1}{i\nu_n-E_X^\mu(\mathbf{k})}-\frac{1}{i\nu_n+E_X^\mu(\mathbf{k})}\right),
\end{align}
and taking into account that when using the contour integration theorem Matsubara sums become:
\begin{align}
    &\frac{1}{\beta}\sum_{in}F(i\nu_n)=\frac{1}{2\pi i}\oint_C\frac{dz}{2\pi i} h(z)F(z)
    \nonumber \\
    &=-\frac{1}{\beta}\sum_{z_0}\text{Res}[F(z_0)]h(z_0),
    \label{Matsubara}
\end{align}
with $C$ being a closed path enclosing $F$'s poles ($z_0$) and $h(z)$ representing the Bose-Einstein or Fermi-Dirac distributions, depending on whether bosons or fermions are being considered, expressions \eqref{eq: ap Qf} and \eqref{eq:Qx A} can be further simplified to: 
\begin{equation}
    Q^f_{ij}=\frac{1}{N}\sum_{\mu,\mathbf{k}}\eta_i^\mu(\mathbf{k})\eta_j^{\mu*}(\mathbf{k})\frac{U}{2E_X^\mu(\mathbf{k})}[b(E_X^\mu(\mathbf{k}))-b(-E_X^\mu(\mathbf{k}))],
\end{equation}
\begin{equation}
    Q^X_{ij}=\frac{1}{N}\sum_{\mu,\mathbf{k},\alpha}e^{-i\mathbf{k}\cdot\mathbf{r}_{ij}}\xi_{i\alpha}^{\mu}(\mathbf{k})\xi_{j\alpha}^{\mu*}(\mathbf{k})f(\epsilon_f^{\mu}(\mathbf{k})),
\end{equation}
which are the two first SRMFT self-consistent equations introduced in the text, \eqref{eq:Qf sc} and \eqref{eq:Qx sc}. Remember that here $f(x)$ and $b(x)$ are the Fermi-Dirac and Bose-Einstein distributions respectively.

Repeating the same procedure for the equation given by the constraint:
\begin{align}
    1&=\frac{1}{N_c}\sum_i\langle X_i^*X_i\rangle_X
    =\frac{1}{N_cN}\sum_{\mu\mathbf{k}}\langle X_\mu^*(\mathbf{k})X_\mu(\mathbf{k})\rangle_X
    \nonumber \\
    &=\frac{1}{N_cN}\sum_{\mu\mathbf{k}}\frac{U}{2E_X^\mu(\mathbf{k})}[b(E_X^\mu(\mathbf{k}))-b(-E_X^\mu(\mathbf{k}))].
\end{align}
the last self-consistent equation \eqref{eq: constraint} is retrieved. Recall that here $N_c$ is the number of sites per unit cell.

Another feature that is important to highlight is that:
\begin{align}
&\langle i\sum_{\alpha,\beta}f^\dagger_{i\alpha}\mathbf{\tau}_{\alpha\beta}\cdot\frac{\mathbf{d}_{il}\times \mathbf{d}_{lj}}{|\mathbf{d}_{il}\times \mathbf{d}_{lj}|}f_{j\beta}\rangle_f \nonumber\\
=&\frac{1}{N}\sum_{\mathbf{k},\alpha,\beta}e^{-i\mathbf{k}\cdot\mathbf{r}_{ij}}\langle if^\dagger_{i\alpha}(\mathbf{k}){\tau}_{\alpha\beta}\cdot\frac{\mathbf{d}_{il}\times \mathbf{d}_{lj}}{|\mathbf{d}_{il}\times \mathbf{d}_{lj}|} f_{j\beta}(\mathbf{k})\rangle_f
\nonumber \\
=&\frac{1}{N}\sum_{\mathbf{k},\mu,\alpha,\beta}e^{-i\mathbf{k}\cdot\mathbf{r}_{ij}}\tilde{\xi}_{i\alpha}^{\mu}(\mathbf{k}) \tilde{\xi}_{j\beta}^{\mu*}(\mathbf{k})\langle f^\dagger_{\mu}(\mathbf{k})f_{\mu}(\mathbf{k})\rangle_f.
\end{align}

As we can see, the spin-dependent factor is fully absorbed by the new eigenvectors, ${\tilde \xi}_{i\alpha}$, of the kinetic part of the rotor Hamiltonian which now includes the $\lambda_{SOI}$ contribution. Consequently, adding extra terms to our Hamiltonian, such as the Fu-Kane-Mele spin-orbit interaction, may define a different 
$Q_{ij}^X$, but it will ultimately lead to exactly the same set of self-consistent equations.

\section{Dependence of $U_c$ with $\lambda_{SOI}$}
\label{ap:dependence}

\begin{figure}[b]
    \centering
    \includegraphics[width=0.9\linewidth]{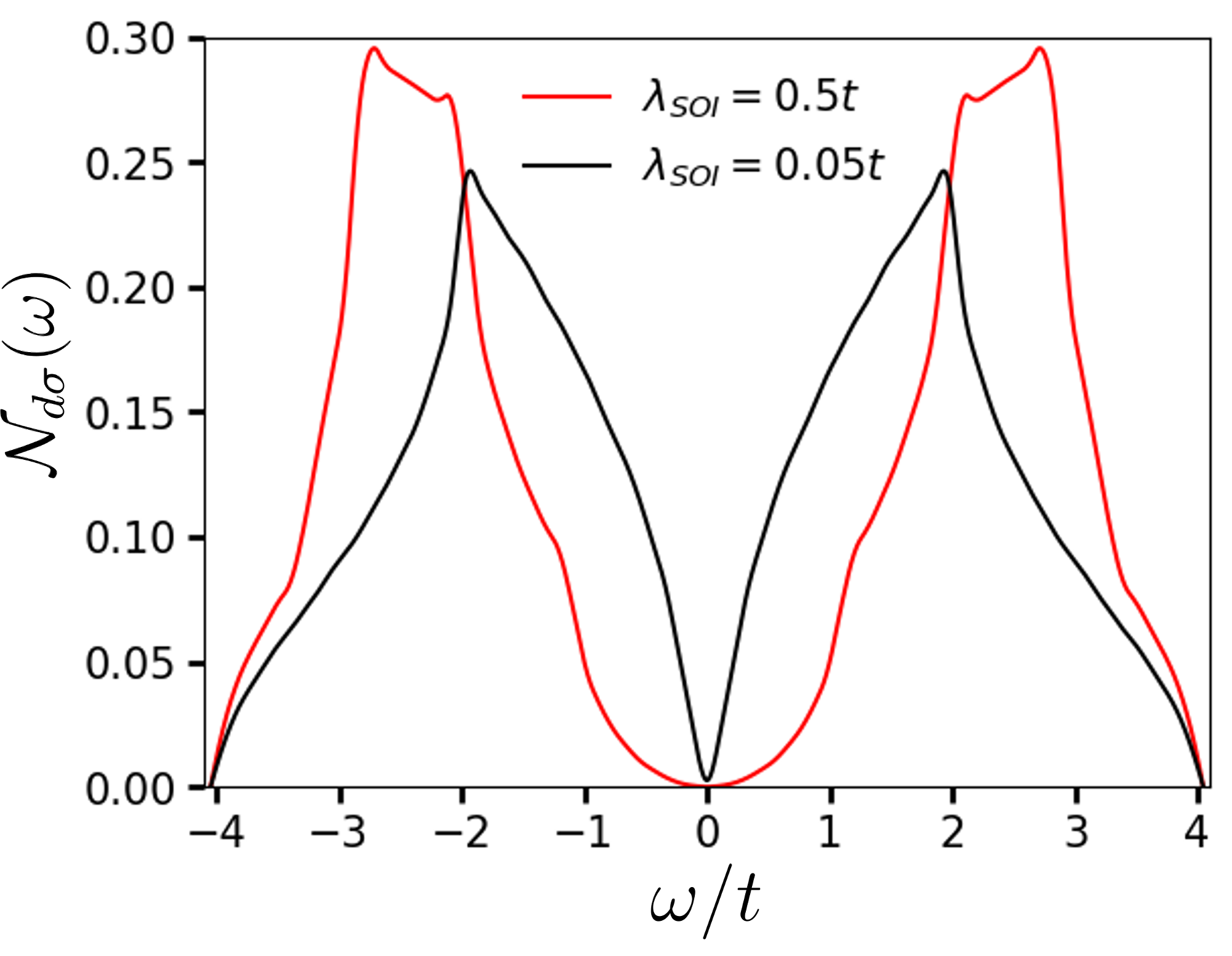}
    \caption{Density of states per spin $\mathcal{N}_{d\sigma}(\omega)$ of the non-interacting Dirac semimetallic Hamiltonian \eqref{eq:Dirac H} for $\lambda_{SOI}/t=0.05$ and $\lambda_{SOI}/t=0.5$.}
    \label{fig:density of states}
\end{figure}

An interesting feature we can observe in Fig. \ref{fig:fig2} of the main text is that $U_c$ decreases with $\lambda_{SOI}$. This can be understood from the $T=0$ expression of $U_c$ \cite{Florens2004}:
\begin{equation}
    \frac{U_c}{U_c^\infty}=\left[\int^D_{-D}d\omega\frac{\mathcal{N}_{d\sigma}(\omega)}{\sqrt{1+\omega/D}}\right]^{-2},\label{eq:Florens}
\end{equation}
where $U_c^\infty=8|\bar{E}|$ (with $\bar{E}$ the average kinetic energy per electron in the non-interacting model)  is the critical Hubbard repulsion at which the Mott transition occurs in the infinite dimensional (strict local) limit, and $\mathcal{N}_{d\sigma}(\omega)$ is the non-interacting density of states per spin, $\mathcal{N}_{d\sigma}(\omega)$, with half-bandwidth $D$. 

In Fig. \ref{fig:density of states}, the density of states $\mathcal{N}_{d\sigma}(\omega)$ of the non-interacting system is shown for increasing $\lambda_{SOI}$. Using \eqref{eq:Florens} we can obtain the $U_c$ dependence on $\lambda_{SOI}$. One finds that $U_c/U_c^\infty$ decreases with $\lambda_{SOI}$ due to enhancement of $\mathcal{N}_{d\sigma}(\omega)$ at high energies with increasing $\lambda_{SOI}$ as observed in Fig. \ref{fig:density of states}. Concomitantly  
 $|\bar{E}|$ (and $U_c^\infty$) increases with $\lambda_{SOI}$ for similar reasons. The net effect is a suppression of $U_c/t$ with $\lambda_{SOI}$ since the enhancement in $\mathcal{N}_{d\sigma}(\omega)$ occurs close to $\pm D$. More precisely, we find $U_c(\lambda_{SOI}=0.5t)/U_c(\lambda_{SOI}=0.05t)=0.69$ in very good agreement with the phase diagram of Fig. \ref{fig:fig2}, for which $U_c(\lambda_{SOI}=0.5t)/U_c(\lambda_{SOI}=0.05t)\sim 0.63$. Our analysis highlights the reliability of Eq. \ref{eq:Florens} for estimating $U_c/t$ even in multiband systems.

From this analysis, we can generally conclude that if $D$ is nearly independent of $\lambda_{SOI}$ as in Fig. \ref{fig:density of states}, the shape of $\mathcal{N}_{d\sigma}(\omega)$ determines the behavior of $U_c$. Systems where $\mathcal{N}_{d\sigma}(\omega)$ exhibits higher density of states toward higher energies close to the band edges would yield lower values of $U_c$. A qualitatively similar relationship between $\lambda_{SOI}$ and $U_c$ was found previously in the context of pyrochlore iridates\cite{Pesin2010}.

\section{Electron Green's function zeros and spectral density}
\label{ap: Green}

In this appendix, we show the explicit derivation of the non-local electron Green’s function, 
$G_{d\sigma}({\bf k} ,\omega)$, for the nodal-line semimetal under the influence of electronic interactions. 

In \eqref{eq:nonlocalGreen} we show the expression of the electron Green's function as the convolution of the spinon and rotor Green's functions.\cite{Sangiovanni2024} The non-local Green's function for the spinons and rotors, respectively, read:
\begin{align}
    G_{f\sigma}^{ss'}({\bf k },i\omega)=[i\omega \mathbb{I}-H_{f\sigma}({\bf k})]^{-1}\\
    G_{X\sigma}^{ss'}({\bf k },i\nu_n)=[(\frac{\nu_n^2}{U}+\rho)\mathbb{I}-H_{X}^{(1)}({\bf k})]^{-1},
\end{align}
where, for $\lambda_{SOI}=0$ and just considering n.n. hoping:
\begin{align}
    H_{f\sigma}({\bf k})=\begin{pmatrix}0 &-Q_fd_{AB}({\bf k})\\
    -Q_fd_{BA}({\bf k}) & 0
    \end{pmatrix},\nonumber\\
    H_{X}({\bf k})=\begin{pmatrix}0 &-Q_Xd_{AB}({\bf k})\\
    -Q_Xd_{BA}({\bf k}) & 0
    \end{pmatrix},    
\end{align}
being $d_{AB}({\bf k})\equiv d_{1}({\bf k})+id_{2}({\bf k})=d_{BA}({\bf k})^*$. The eigenvalues of the spinon and rotor Hamiltonians are $\epsilon_f^{\pm}({\bf k})$ and $\epsilon_X^{\pm}({\bf k})$. Thus, considering this in \eqref{eq:nonlocalGreen} and performing the Matsubara sums using \eqref{Matsubara}, one can get the electron Green's function matrix:
\begin{align}   G_{d\sigma}=\begin{pmatrix}G_{d\sigma}^{ss}&G_{d\sigma}^{AB}\\
    G_{d\sigma}^{AB*} &  G_{d\sigma}^{ss}
    \end{pmatrix},
\end{align}
in which the matrix elements in the insulating phase ($Z=0$) read:
\widetext
\small
\begin{align}  
G_{d\sigma}^{ss}({\bf k},i\omega) &=\frac{1}{4N}\sum_{\bf {\bf q}}\left\{\frac{U}{2E_{X}^+({\bf q})}\left(\frac{f(\epsilon_{f}^+({\bf k}-{\bf q}))+b(-E_{X}^+({\bf q}))}{i\omega-\epsilon_{f}^-({\bf k}-{\bf q})+E_{X}^+({\bf q})}-\frac{f(\epsilon_{f}^+({\bf k}-{\bf q}))+b(E_{X}^+({\bf q}))}{i\omega-\epsilon_{f}^-({\bf k}-{\bf q})-E_{X}^+({\bf q})}\right)\right.\nonumber\\
&+\frac{U}{2E_{X}^+({\bf q})}\left(\frac{f(-\epsilon_{f}^+({\bf k}-{\bf q}))+b(-E_{X}^+({\bf q}))}{i\omega-\epsilon_{f}^+({\bf k}-{\bf q})+E_{X}^+({\bf q})}-\frac{f(-\epsilon_{f}^+({\bf k}-{\bf q}))+b(E_{X}^+({\bf q}))}{i\omega-\epsilon_{f}^+({\bf k}-{\bf q})-E_{X}^+({\bf q})}\right)
\nonumber\\
&+\frac{U}{2E_{X}^-({\bf q})}\left(\frac{f(\epsilon_{f}^+({\bf k}-{\bf q}))+b(-E_{X}^-({\bf q}))}{i\omega-\epsilon_{f}^-({\bf k}-{\bf q})+E_{X}^-({\bf q})}-\frac{f(\epsilon_{f}^+({\bf k}-{\bf q}))+b(E_{X}^-({\bf q}))}{i\omega-\epsilon_{f}^-({\bf k}-{\bf q})-E_{X}^-({\bf q})}\right)\nonumber\\
&\left.+\frac{U}{2E_{X}^-({\bf q})}\left(\frac{f(-\epsilon_{f}^+({\bf k}-{\bf q}))+b(-E_{X}^-({\bf q}))}{i\omega-\epsilon_{f}^+({\bf k}-{\bf q})+E_{X}^-({\bf q})} -\frac{f(-\epsilon_{f}^+({\bf k}-{\bf q}))+b(E_{X}^-({\bf q}))}{i\omega-\epsilon_{f}^+({\bf k}-{\bf q})-E_{X}^-({\bf q})}\right)\right\},
\end{align}
\begin{align}
G_{d\sigma}^{AB}({\bf k},i\omega) &=\frac{1}{4N}\sum_{\bf {\bf q}}c_{{\bf k}{\bf q}}\left\{\frac{U}{2E_{X}^+({\bf q})}\left(\frac{f(\epsilon_{f}^+({\bf k}-{\bf q}))+b(-E_{X}^+({\bf q}))}{i\omega-\epsilon_{f}^-({\bf k}-{\bf q})+E_{X}^+({\bf q})}-\frac{f(\epsilon_{f}^+({\bf k}-{\bf q}))+b(E_{X}^+({\bf q}))}{i\omega-\epsilon_{f}^-({\bf k}-{\bf q})-E_{X}^+({\bf q})}\right)\right.\nonumber\\
&-\frac{U}{2E_{X}^+({\bf q})}\left(\frac{f(-\epsilon_{f}^+({\bf k}-{\bf q}))+b(-E_{X}^+({\bf q}))}{i\omega-\epsilon_{f}^+({\bf k}-{\bf q})+E_{X}^+({\bf q})}-\frac{f(-\epsilon_{f}^+({\bf k}-{\bf q}))+b(E_{X}^+({\bf q}))}{i\omega-\epsilon_{f}^+({\bf k}-{\bf q})-E_{X}^+({\bf q})}\right)
\nonumber\\
&-\frac{U}{2E_{X}^-({\bf q})}\left(\frac{f(\epsilon_{f}^+({\bf k}-{\bf q}))+b(-E_{X}^-({\bf q}))}{i\omega-\epsilon_{f}^-({\bf k}-{\bf q})+E_{X}^-({\bf q})}-\frac{f(\epsilon_{f}^+({\bf k}-{\bf q}))+b(E_{X}^-({\bf q}))}{i\omega-\epsilon_{f}^-({\bf k}-{\bf q})-E_{X}^-({\bf q})}\right)\nonumber\\
&\left.+\frac{U}{2E_{X}^-({\bf q})}\left(\frac{f(-\epsilon_{f}^+({\bf k}-{\bf q}))+b(-E_{X}^-({\bf q}))}{i\omega-\epsilon_{f}^+({\bf k}-{\bf q})+E_{X}^-({\bf q})} -\frac{f(-\epsilon_{f}^+({\bf k}-{\bf q}))+b(E_{X}^-({\bf q}))}{i\omega-\epsilon_{f}^+({\bf k}-{\bf q})-E_{X}^-({\bf q})}\right)\right\},
\end{align}
\twocolumngrid
where: $E_X^\pm({\bf q})=\sqrt{U(\rho+ \epsilon_X^{\pm}({\bf q}))}$ and $c_{{\bf k}{\bf q}}\equiv \frac{d_{AB}({\bf q})d_{AB}({\bf k}-{\bf q})}{|d_{AB}({\bf q})||d_{AB}({\bf k}-{\bf q})|}$.

Since $Z\neq 0$ in the metallic phase, the coherent part of the Green's function, $G_{d\sigma}^{coh}({\bf k}, i\omega)$ plays a role, and reads:
\begin{align}
    G_{d\sigma}^{ss,coh}({\bf k},i\omega)=&\frac{Z}{2}\left(\frac{1}{i\omega -\epsilon_f^+({\bf k})}+\frac{1}{i\omega -\epsilon_f^-({\bf k})}\right) \nonumber \\
    G_{d\sigma}^{AB,coh}({\bf k},i\omega)=&\frac{Z}{2}\frac{d_{AB}{\bf k}}{|d_{AB}({\bf k})|}\left(\frac{1}{i\omega -\epsilon_f^+({\bf k})}-\frac{1}{i\omega -\epsilon_f^-({\bf k})}\right). 
    \nonumber \\ \nonumber
\end{align}
Thus, performing the analytical continuation $i\omega\rightarrow \omega+i0^+$, one is able to compute the determinant of the non-local electron Green's function shown in Fig. \ref{fig:Green_spectral}. The spectral density function, $A_{d\sigma}({\bf k},\omega)=-\frac{1}{\pi}\operatorname{\mathbb{I}m}[G^{loc}_{d\sigma}({\bf k},\omega+i0^+)]$, which instead involves the local Green's function, $G^{loc}_{d\sigma}$,\cite{Manuel2022,Manuel2024} is also shown.
\bibliography{TFM}
\end{document}